\documentclass[11pt]{article}

\usepackage[final]{acl}

\usepackage{times}
\usepackage{latexsym}

\usepackage[T1]{fontenc}

\usepackage[utf8]{inputenc}

\usepackage{microtype}

\usepackage{inconsolata}

\usepackage{graphicx}
\usepackage{hyperref}
\usepackage{url}
\usepackage{paralist}
\usepackage{graphicx}
\usepackage{wrapfig}
\usepackage{booktabs}
\usepackage{multirow}
\usepackage{amsmath}
\usepackage{cleveref}
\usepackage{amssymb}
\usepackage{ amsfonts}
\usepackage{algorithm}
\usepackage{algorithmic}
\usepackage{subcaption}
\usepackage{todonotes}
\usepackage{enumitem}
\usepackage[utf8]{inputenc}
\usepackage{titletoc}   
\usepackage{bm} 
\usepackage[T2A,T1]{fontenc} 
\newcommand{\textcyr}[1]{{\fontencoding{T2A}\selectfont #1}} 
\newcommand{\vect}[1]{\boldsymbol{#1}}

\usepackage{booktabs}  
\usepackage{caption}   
\usepackage[table]{xcolor}

\definecolor{Mycolor1}{HTML}{BAD8F2}
\definecolor{Mycolor2}{HTML}{E0F0FA}
\definecolor{reda}{RGB}{192,0,0}
\definecolor{color2}{RGB}{0,128,0}
\definecolor{mypink}{rgb}{1.0, 0.85, 0.85} 
\definecolor{mygreen}{rgb}{0.1, 0.5, 0.1} 
\definecolor{myred}{rgb}{0.7, 0.1, 0.1}   
\usepackage{fontawesome5} 
\usepackage{seqsplit}

\newenvironment{itemize*}%
 {\leftmargini=10pt\begin{compactitem}%
  \setlength{\itemsep}{0pt}%
  \setlength{\parskip}{0pt}%
  }%
 {\end{compactitem}}
 
\newenvironment{enumerate*}%
 {\begin{enumerate}%
  \setlength{\itemsep}{0pt}%
  \setlength{\parskip}{0pt}}%
 {\end{enumerate}}

\title{Vocabulary Hijacking in LVLMs: Unveiling Critical Attention Heads by Excluding Inert Tokens to Mitigate Hallucination}

\author{
    Yangneng Chen$^1$
    \quad Junlin Li$^1$
    \quad Weijun Yao$^2$
    \quad Xilai Ma$^1$\\
    \quad \textbf{Guodong Du}$^3$
    \quad \textbf{Wenya Wang}$^4$
    \quad  \textbf{Jing Li}$^1$\textsuperscript{\texorpdfstring{\faIcon[regular]{envelope}}{}} 
    \\$^{1}$Harbin Institute of Technology (Shenzhen), China  \quad $^{2}$Huawei Technologies Co., Ltd.   
    \\$^{3}$The Hong Kong Polytechnic University \quad
     $^{4}$
Nanyang Technological University\\
    \texttt{yangnengchen8@gmail.com} \quad \texttt{jingli.phd@hotmail.com}  
}

\begin{document}
\maketitle
\begin{abstract}
Large Vision-Language Models (LVLMs) have achieved remarkable progress in multimodal tasks, yet their reliability is persistently undermined by hallucinations—generating text that contradicts visual input. 
Recent studies often attribute these errors to inadequate visual attention. In this work, we analyze the attention mechanisms via the logit lens, uncovering a distinct anomaly we term \textbf{Vocabulary Hijacking}. 
We discover that specific visual tokens, defined as \textbf{Inert Tokens}, disproportionately attract attention. 
Crucially, when their intermediate hidden states are projected into the vocabulary space, they consistently decode to a fixed set of unrelated words (termed \textbf{Hijacking Anchors}) across layers, revealing a rigid semantic collapse.
Leveraging this semantic rigidity, we propose \textbf{Hijacking Anchor-Based Identification (HABI)}, a robust strategy to accurately localize these Inert Tokens. 
To quantify the impact of this phenomenon, we introduce the \textbf{Non-Hijacked Visual Attention Ratio (NHAR)}, a novel metric designed to identify attention heads that remain resilient to hijacking and are critical for factual accuracy. 
Building on these insights, we propose \textbf{Hijacking-Aware Visual Attention Enhancement (HAVAE)}, a training-free intervention that selectively strengthens the focus of these identified heads on salient visual content. 
Extensive experiments across multiple benchmarks demonstrate that HAVAE significantly mitigates hallucinations with \textbf{no additional computational overhead}, while preserving the model's general capabilities. Our code is publicly available at \url{https://github.com/lab-klc/HAVAE}.
\let\thefootnote\relax\footnotetext{\faIcon[regular]{envelope}~Corresponding author.}
\end{abstract}

\section{Introduction}

Large Vision-Language Models (LVLMs), extending large language models (LLMs) to process visual inputs, represent a major advancement in artificial intelligence~\citep{liu2023llava,shikra,minigpt4,wang2024qwen2}. 
Nevertheless, LVLMs remain susceptible to hallucinations~\citep{llavarlhf, lure, huang2023opera, bai2024hallucination}, where generated text fails to align with visual content. 
Such inconsistencies undermine LVLM reliability in multimodal tasks, limiting their practical deployment.

\begin{figure}[t]
    \centering
    \includegraphics[width=\linewidth]{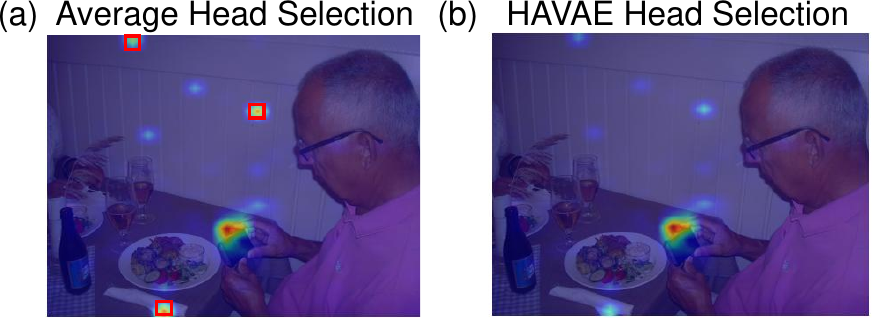}
    \caption{Attention maps for generating ``phone''. (a) The average of all heads exhibits the Vocabulary Hijacking phenomenon, with focus scattered on the background. (b) In contrast, heads selected by HAVAE concentrate attention precisely on the target object.}
    \vspace{-4mm}
    \label{fig:head_intro}
\end{figure}

Substantial research on mitigating hallucinations LVLMs has produced various intervention strategies~\citep{2023rlhf-v, VCD, chen2024halc, lihidden}. 
\textbf{A critical consensus has emerged}: insufficient attention to visual tokens during generation is a primary driver of hallucinations.
This insight has inspired numerous attention-based methods~\citep{jiang2025devils,liu2024paying,yang2025mitigating}.
However, a fundamental question remains unresolved: \textit{Which specific attention heads are actually critical?}

\begin{figure*}[t]
    \centering
    \includegraphics[width=\linewidth]{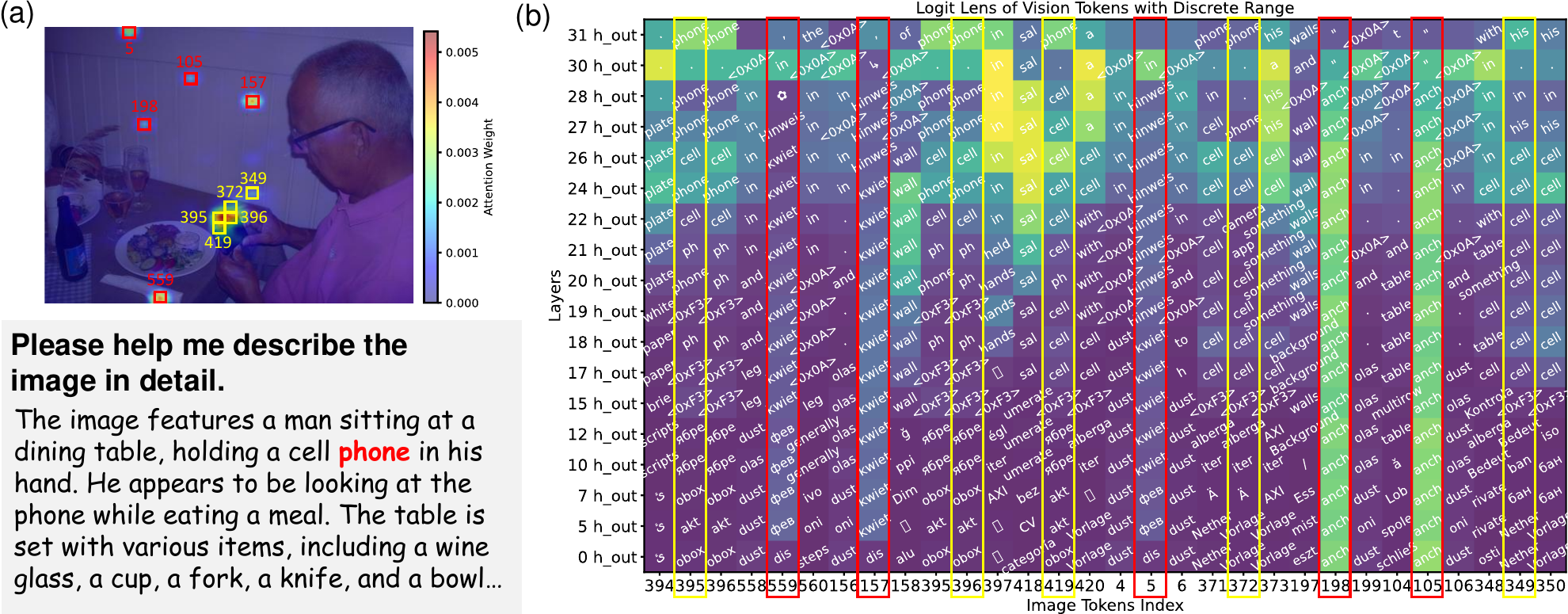}
    \caption{
    Illustration of the Vocabulary Hijacking phenomenon in LLaVA-1.5 7B. 
    (a) Attention heatmap for the token ``phone'', where several irrelevant background tokens capture excessive attention (highlighted in red boxes), which we term \textbf{Inert Tokens}.
    (b) Logit lens analysis of the top-10 attended visual tokens and their neighbors. 
\textbf{Inert Tokens} (red boxes) exhibit rigid non-semantic traces, contrasting sharply with target tokens (yellow boxes) that evolve from meaningless words into image-aligned concepts.
    }
    \vspace{-4mm}
    \label{fig:intro}
\end{figure*}
To demystify attention mechanisms, \citep{jiang2025devils} applied Logit Lens to vision tokens, identifying a transition from \textit{Visual Information Enrichment} to \textit{Semantic Refinement}. 
Adopting this framework, we define the \textbf{Trace} as the layer-wise sequence of vocabulary words decoded from a token's hidden states.
Typically, a functional Trace evolves from meaninglessness to rich semantics; for instance, Token \#395 in \Cref{fig:intro} shifts from nonsensical outputs to image-aligned concepts like ``phone'' after layer 20.

Further investigating the top-attended tokens reveals a critical anomaly: \textbf{LVLMs exhibit a tendency to persistently allocate excessive attention to certain background image tokens, whose Traces are dominated by a small set of fixed, meaningless words.}
For example, in \Cref{fig:intro}, tokens \#595 and \#157 receive high attention despite being irrelevant to the generated ``phone''. 
Their Traces remain monopolized by ``kwiet'', failing to exhibit the standard semantic evolution.

We term this phenomenon \textbf{Vocabulary Hijacking}, wherein specific vision tokens, designated as \textbf{Inert Tokens}, exhibit \textit{inertia} across layers while attracting substantial attention.
To characterize this, we introduce the \textbf{Vocabulary Hijacking Score (VHS)}. Statistical analysis reveals that the Traces of Inert Tokens are dominated by a small subset of words with anomalously high VHS (e.g., ``kwiet''), which we formally define as \textbf{Hijacking Anchors}.

Leveraging this finding, we propose \textbf{Hijacking Anchor-Based Identification (HABI)}. By quantifying the proportion of a token's Trace that coincides with Hijacking Anchors, HABI effectively localizes Inert Tokens.
While Vocabulary Hijacking superficially resembles the \textit{Visual Attention Sink}~\citep{visattnsink}, empirical results demonstrate significant divergences (detailed in Appendix~\ref{app:vas_comparison}), confirming that our framework captures this anomalous attention phenomenon more precisely and robustly.

Building upon this foundation, we investigate the intrinsic link between Vocabulary Hijacking and hallucinations. 
Statistical experiments reveal a strong positive correlation: excessive attention to Inert Tokens significantly increases the propensity for hallucination.
Motivated by this, we propose \textbf{Non-Hijacked Visual Attention Ratio (NHAR)} to identify critical heads focusing on salient visual information, and introduce \textbf{Hijacking-Aware Visual Attention Enhancement (HAVAE)} to selectively amplify them. 
As shown in \Cref{fig:head_intro}, HAVAE-selected heads exhibit precise grounding on target objects, contrasting sharply with the scattered attention of generic heads.

Extensive experiments validate HAVAE's superiority: it achieves SOTA performance across hallucination benchmarks with \textbf{zero additional computational overhead}, while fully preserving general capabilities. 
This highlights an optimal balance between performance and practical efficiency.
Our main contributions are summarized as follows:
\begin{itemize}
    \item We unveil the novel mechanism of \textbf{Vocabulary Hijacking} and propose \textbf{HABI}, which reliably localizes \textbf{Inert Tokens} across various LVLMs.
    \item We introduce \textbf{HAVAE}, a training-free intervention that leverages our novel metric, \textbf{NHAR}, to identify and selectively reinforce the attention heads most critical for hallucination mitigation.
    \item Through extensive experimentation across diverse architectures, we demonstrate that HAVAE establishes a new SOTA in hallucination mitigation with zero computational overhead, while preserving general capabilities.
\end{itemize}

\section{Related Work}
\label{sec:related}
\paragraph{Reducing Hallucinations in MLLMs.} 
Mitigating hallucinations in MLLMs is primarily addressed through fine-tuning~\citep{2023rlhf-v,yu2024rlaifv} or training-free inference interventions. 
Prominent strategies include contrastive decoding~\citep{VCD, chen2024halc}, activation steering~\citep{lihidden}, and attention intervention~\citep{liu2024paying, jiang2025devils, yang2025mitigating}. 
While attention-based methods are most relevant to our work, they face a critical unresolved problem: \textbf{determining precisely \textit{which} heads require intervention.} 
In the absence of a principled identification mechanism, these approaches are often forced to rely on heuristics.
Our work bridges this gap by proposing a precise metric to target the specific heads essential for factual grounding.

\paragraph{Attention Sink in Language Models.}
The ``Attention Sink'' phenomenon, where vacuous tokens monopolize attention, is established in LLMs~\citep{xiao23streamingllm} and recently extended to LVLMs as Visual Attention Sinks (VAS)~\citep{visattnsink}. 
Along similar lines, works like AVISC~\citep{AVISC} have also investigated comparable attention anomalies to mitigate hallucinations. 
However, the underlying mechanisms of these phenomena remain under-explored. 
We argue that our proposed \textit{Vocabulary Hijacking} offers a more precise characterization of anomalous attention patterns in LVLMs compared to the VAS framework.

\section{Characterizing and Identifying Vocabulary Hijacking}
\subsection{Preliminaries}

\paragraph{Autoregressive Generation in LVLMs.}
Large Vision-Language Models (LVLMs) generate responses autoregressively by modeling the conditional probability of the next token. 
At each timestep $k$, the model predicts token $y_k$ based on the preceding context, which comprises a sequence of image tokens $\mathcal{I}_\text{v}$, a text prompt $\mathcal{I}_\text{t}$, and previously generated tokens $\mathcal{I}_\text{o}$. 
These components are concatenated to form a single input sequence $\mathcal{I}$.

\noindent\textbf{Attention Mechanism in LVLMs.}
The core component enabling token interaction is Multi-Head Attention (MHA). Following \citep{elhage2021mathematical}, at layer $\ell$, the representation for a token $\vect{x}^{\ell-1}_i$ is updated by attending to all previous tokens $\vect{X}_{\leq i}^{\ell-1} = \{ \vect{x}^{\ell-1}_0, \dots, \vect{x}^{\ell-1}_i \}$ as follows:
\begin{equation} \label{eq:attn}
\begin{split}
    \text{MHA}^{\ell, h} (\vect{x}^{\ell-1}_i) &= \sum_{j \leq i} \vect{A}^{\ell, h}_{i, j} \vect{x}_j^{\ell-1} \vect{W}_{OV}^{\ell, h}, \\
    \vect{A}^{\ell, h}_i &= \text{softmax} \Bigg( \frac{1}{\sqrt{D_k}} (\vect{x}^{\ell-1}_i \vect{W}^{\ell, h}_Q) \\
    &\quad \cdot (\vect{X}^{\ell-1}_{\leq i} \vect{W}^{\ell, h}_K)^\top \Bigg).
\end{split}
\end{equation}

Here, $\vect{W}^{\ell, h}_Q, \vect{W}^{\ell, h}_K \in \mathbb{R}^{D \times D_k}$ are the query and key projection matrices, and $\vect{W}_{OV}^{\ell, h} \in \mathbb{R}^{D \times D}$ is the output-value projection matrix. 
The attention weight $\vect{A}^{\ell, h}_{i, j}$ quantifies the contribution of token $\vect{x}_j^{\ell-1}$ to the updated representation of token $\vect{x}_i^{\ell-1}$. 
Our analysis focuses on the cases where $i \in \mathcal{I}_\text{o}$ and $j \in \mathcal{I}_\text{v}$, which correspond to the attention paid to visual tokens during the generation process.


\noindent\textbf{Logit Lens.}
To investigate the model's internal processing of visual information, we employ Logit Lens~\citep{logitLens,jiang2025devils}.
It projects an intermediate visual hidden state $\vect{v}_i^\ell$ directly onto the vocabulary distribution $\Sigma$ by applying the model's final unembedding matrix, $W_\Sigma \in \mathbb{R}^{|\Sigma| \times d}$:
\begin{equation}
    \mathbf{p}(\Sigma|\vect{v}_i^\ell) = \text{softmax}(\vect{W}_{\Sigma} \cdot \vect{v}_i^\ell) \in \mathbb{R}^{|\Sigma|},
    \label{eq:logit_lens}
\end{equation}
where $p_j(\Sigma|\vect{v}_i^\ell)$ denotes the probability of the $j$-th vocabulary token.
We identify the token with the highest probability from this distribution as the semantic interpretation of the hidden state $\vect{v}_i^\ell$.

\subsection{Setup for Exploratory Analysis}
\label{sec:case_setup}

Following \citep{jiang2025devils}, we conduct our analysis on 500 random images from the COCO 2014 validation set~\citep{lin2014mscoco}, utilizing LLaVA-1.5 (7B/13B), Shikra-7B, MiniGPT4-7B, and Qwen2-VL-7B. 
Descriptions are generated via \textbf{greedy decoding} with the prompt: \emph{``Please help me describe the image in detail''}. 
Let $\mathcal{V}$ denote the collective set of vision tokens from these images.
We categorize generated objects into real ($\mathcal{O}_{\mathrm{real}}$) and hallucinated ($\mathcal{O}_{\mathrm{hal}}$) sets based on ground-truth annotations of COCO.

\begin{figure*}[t]
    \centering
    \includegraphics[width=0.95\linewidth]{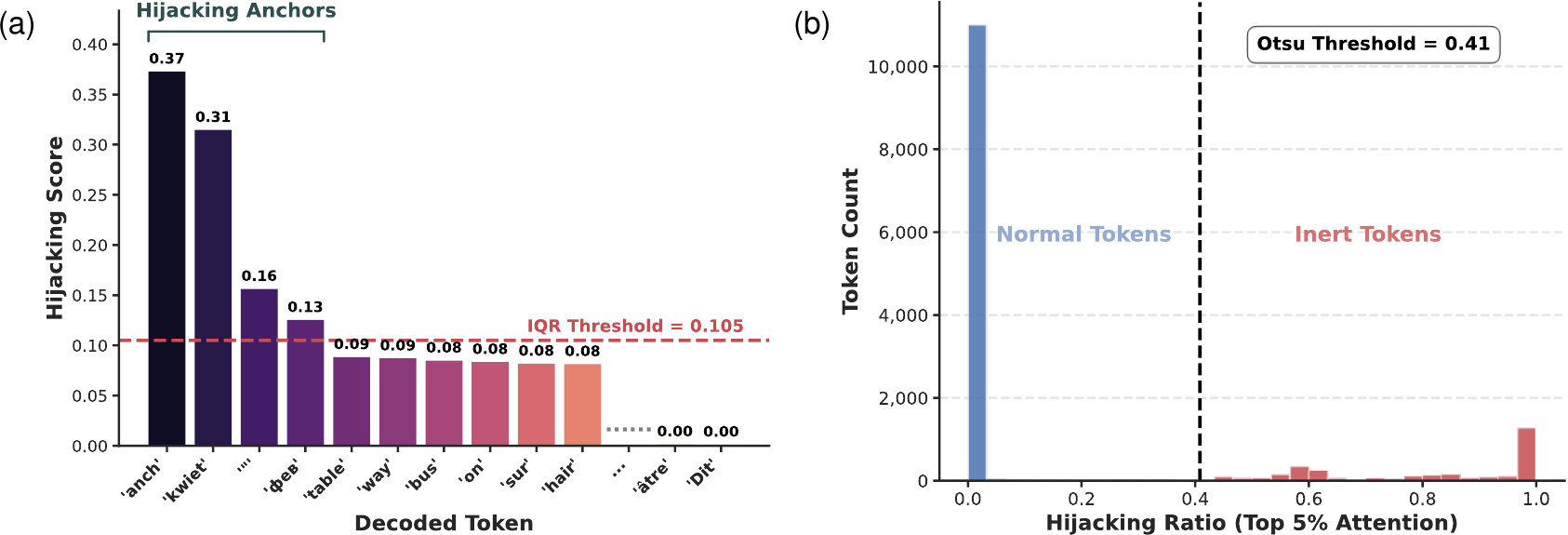}
    \caption{
    Empirical basis for the HABI method on LLaVA-1.5 7B. 
    \textbf{(a)} The long-tailed distribution of Mean Hijacking Scores, illustrating the identification of \textbf{Hijacking Anchors} via an IQR-based outlier threshold. 
    \textbf{(b)} The distinct bimodal distribution of the Hijacking Ratio for salient visual tokens (top 5\% attention mass), demonstrating the automatic isolation of \textbf{Inert Tokens} using Otsu's method.
    }
    \label{fig:threshod}
\end{figure*}

\subsection{Understanding Vocabulary Hijacking via Logit Lens}
\label{sec:vf}

\paragraph{Trace Analysis \& Hijacking Score.} 
For a vision token $v_i \in \mathcal{V}$, we define its \textbf{Trace}, $\mathcal{T}_{v_i}$, as the sequence of words decoded from its layer-wise hidden states via Logit Lens.
We identify \textbf{Inert Tokens} as specific high-attention tokens that are rigidly dominated by fixed, meaningless words.
To identify them, we designate the most frequent token in $\mathcal{T}_{v_i}$ as the \textbf{Anchor} ($a_{v_i}$) and propose a composite \textbf{Hijacking Score} based on three dimensions:
(1) \textbf{Dominance ($\mathcal{D}$)}: the proportion of the Anchor $a_{v_i}$ within the Trace $\mathcal{T}_{v_i}$; 
(2) \textbf{Frequency ($\mathcal{F}$)}: the log-smoothed frequency of $a_{v_i}$ among all Anchors in $\mathcal{V}$; and 
(3) \textbf{Attention ($\mathcal{A}$)}: the log-smoothed average attention received by $v_i$ during the first 10 decoding steps.
The score is formulated as:
\begin{equation}
    S_{\text{hijack}}(v_i) = \mathcal{D}(v_i) \cdot \mathcal{F}(v_i) \cdot \mathcal{A}(v_i).
\end{equation}

\paragraph{Defining Hijacking Anchors.}
We shift focus to the vocabulary level $\Sigma$ to identify systemic distractors. 
For a unique token $w \in \Sigma$, we compute its \textbf{Mean Hijacking Score}, $\bar{S}(w)$, by averaging $S_{\text{hijack}}$ across all vision tokens anchored to it:
\begin{equation}
    \bar{S}(w) = \frac{1}{|\{v \in \mathcal{V} \mid a_v = w\}|} \sum_{v \in \mathcal{V}, a_v = w} S_{\text{hijack}}(v).
\end{equation}
The distribution of $\bar{S}(w)$ exhibits a long-tailed pattern (Figure~\ref{fig:threshod} (a)). 
Using an outlier threshold $\tau_s = Q_3 + 1.5 \cdot \text{IQR}$, we formally define the set of \textbf{Hijacking Anchors}, $\mathcal{A}_{\text{hijack}}$, as:
\begin{equation}
    \mathcal{A}_{\text{hijack}} = \{ w \in \Sigma \mid \bar{S}(w) > \tau_s \}.
\end{equation}
This set characterizes a systemic model property, indicating that vision tokens whose Traces contain these Anchors are significantly more prone to hijacking the model's attention. 
To further corroborate the validity of the Hijacking Score and the rationality of the IQR thresholding, we present a supplementary clustering analysis in \Cref{app:cluster_analysis}.

\subsection{Hijacking Anchor-Based Identification}
Leveraging the identified set $\mathcal{A}_{\text{hijack}}$, we propose \textit{Hijacking Anchor-Based Identification} (HABI) to detect Inert Tokens. 
We first define the \textbf{Hijacking Ratio} ($r_{\text{hijack}}$) to quantify the extent to which a token's Trace is dominated by Hijacking Anchors:
\begin{equation}
    r_{\text{hijack}}(v_i) = \frac{1}{|\mathcal{T}_{v_i}|} \sum_{t \in \mathcal{T}_{v_i}} \mathbb{I}(t \in \mathcal{A}_{\text{hijack}}).
\end{equation}

To establish a robust decision boundary, we analyze the distribution of $r_{\text{hijack}}$. 
While the global distribution is long-tailed, restricting analysis to salient tokens (top 5\% attention mass, given the sparsity of visual attention) reveals a distinct \textbf{bimodal structure} (Figure~\ref{fig:threshod} (b)).
This bimodality enables the use of \textbf{Otsu's method}~\citep{otsu1979threshold} to automatically derive an optimal separation threshold $\tau_r$. 
Accordingly, we formally define the set of \textbf{Inert Tokens}, $\mathcal{I}_{\text{inert}}$, as the subset of vision tokens exceeding this threshold:
\begin{equation}
    \mathcal{I}_{\text{inert}} = \{ v_i \in \mathcal{I}_\text{v} \mid r_{\text{hijack}}(v_i) > \tau_r \}.
\end{equation}
This definition explicitly isolates the specific subset of tokens responsible for hijacking the model's attention mechanism. 
The specific algorithmic procedure is detailed in \Cref{alg:habi}.
Distributions of Hijacking Scores and Hijacking Ratios for additional models are presented in \Cref{app:HABI}.
Furthermore, we investigate the spatial distribution of Inert Tokens in \Cref{app:VAS_idnex} and provide an attention analysis in \Cref{app:inert_attention_analysis}.
\begin{figure*}[t]
    \centering
    \includegraphics[width=0.95\linewidth]{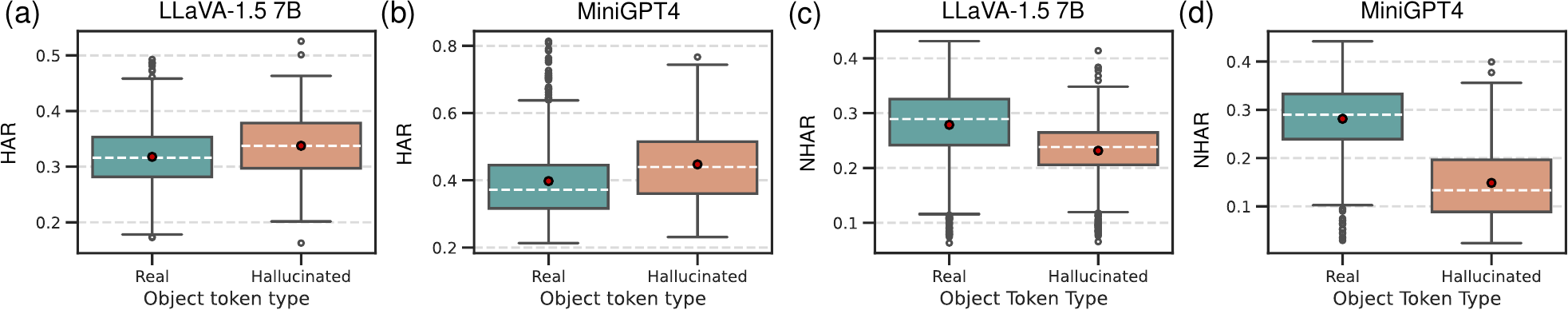}
    \caption{
    Statistical distributions of attention metrics. 
    \textbf{(a, b)} present the distributions of the Hijacked Attention Ratio (HAR) for LLaVA-1.5 7B and MiniGPT-4, respectively. 
    \textbf{(c, d)} illustrate the Non-Hijacked Visual Attention Ratio (NHAR) distributions for the same models.
    }
    \label{fig:HAR_NHAR}
\end{figure*}

\section{From Vocabulary Hijacking to Hallucination}

\subsection{Relationship with Visual Attention Sink}
\label{sec:relation_vas}

As discussed in \Cref{sec:related}, the \textit{Visual Attention Sink} (VAS)~\citep{visattnsink} is characterized by three key features: persistent attention attraction, background localization, and negligible contribution. 
While Vocabulary Hijacking shares these fundamental properties, we demonstrate in \Cref{app:vas_comparison} that our framework offers a significantly more precise mechanism for identifying and characterizing these anomalous attention patterns.

\subsection{Linking Vocabulary Hijacking to Hallucination}
\label{sec:correlation}

\paragraph{Diagnostic Analysis using HAR.}
Acknowledging the functional specialization of attention heads~\citep{Deiseroth2023atman, zhang2024pasta, ge2024model, zheng2024attentionheadsurvey}, we first focus our analysis on the top-300 heads with the \textbf{highest total visual attention}. 
We identify this subset as the primary conduits for visual processing, serving as the basis to probe the hijack-hallucination link.
We introduce the \textbf{Hijacked Attention Ratio (HAR)} to quantify severity of hijacking within this critical visual stream:
\begin{equation}
    \text{HAR}^{(\ell,h)} \triangleq \frac{\sum_{v \in \mathcal{I}_{\text{inert}}} A^{(\ell,h)}_{v}}{\sum_{v \in \mathcal{I}_v} A^{(\ell,h)}_{v}}.
\end{equation}
Our results reveal a strong correlation: \textbf{hallucinated tokens consistently exhibit higher HAR} (Figure~\ref{fig:HAR_NHAR} (a, b)), indicating that when these critical visual heads are ``hijacked,'' the resulting loss of grounding precipitates hallucinations.

\paragraph{From Diagnosis to Intervention: The NHAR Metric.}
While HAR diagnoses distraction, selecting effective heads requires a dual filter: ensuring high visual intensity \textit{and} low hijacking.
To identify heads that actively encode meaningful features using a single standard, we propose the \textbf{Non-Hijacked Visual Attention Ratio (NHAR)}:
\begin{equation}
    \text{NHAR}^{(\ell,h)}(y_k) \triangleq \sum_{v \in \mathcal{I}_v \setminus \mathcal{I}_{\text{inert}}} A^{(\ell,h)}_{k, v}.
    \label{eq:nvar}
\end{equation}
Unlike HAR, NHAR acts as a \textbf{unified density metric}, directly measuring the global attention budget allocated to valid visual content.
Validating on the top-450 NHAR-ranked heads reveals a stark contrast: \textbf{Real Object Tokens correlate with high NHAR, whereas Hallucinated Tokens cluster in low-NHAR regions} (Figure~\ref{fig:HAR_NHAR} (c, d)).
This confirms NHAR as a robust discriminator for factual grounding, justifying its role as our primary selection criterion.
Additional statistics for other models are provided in Appendix~\ref{app:nvar_vasr}.

\begin{table*}[htbp!]
\centering
\resizebox{0.9\textwidth}{!}{%
\begin{tabular}{ll|ccc|cc|cc}
\toprule
\multicolumn{1}{l}{\multirow{2}{*}{\textbf{Model}}}
&\multicolumn{1}{l}{\multirow{2}{*}{\textbf{Method}}}
& \multicolumn{3}{c}{\textbf{CHAIR}} 
& \multicolumn{2}{c}{\textbf{POPE}} 
& \multicolumn{2}{c}{\textbf{POPE Chat}} \\ 
\cmidrule(lr){3-5} \cmidrule(lr){6-7} \cmidrule(lr){8-9}
& & \textbf{CHAIR$_s$ $\downarrow$} & \textbf{CHAIR$_i$ $\downarrow$}& \textbf{F1 $\uparrow$} & \textbf{Acc. $\uparrow$} & \textbf{F1 $\uparrow$} & \textbf{Acc. $\uparrow$} & \textbf{F1 $\uparrow$} \\
\midrule
\multicolumn{1}{l}{\multirow{5}{*}{LLaVA-1.5-7B}} & Greedy        & 48.2 & 14.2 & \cellcolor{gray!15} 76.4 & 84.8 & 85.5 & 85.5 & 83.4 \\
& PAI               & 23.8 & 6.2  &\cellcolor{gray!15} 76.8 & 85.9 & 86.0 & 85.5 & 83.4 \\
& Devils            & 27.2 & 7.0  &\cellcolor{gray!15} 76.1 & 85.5 & 85.8 & 87.6 & 86.9 \\
& VISTA   & 15.6 & 5.2 & \cellcolor{mypink}67.3 & 83.1 & 84.6 & --- & --- \\
& \textbf{HAVAE(Ours)} & \textbf{18.2}$^{\textcolor{mygreen}{\scriptsize -23.5\%}}$ & \textbf{3.8}$^{\textcolor{mygreen}{\scriptsize -38.7\%}}$  &\cellcolor{gray!15} 76.7 & \textbf{86.2}$^{\textcolor{mygreen}{\scriptsize +0.3\%}}$ & \textbf{86.3}$^{\textcolor{mygreen}{\scriptsize +0.3\%}}$ & \textbf{88.0}$^{\textcolor{mygreen}{\scriptsize +0.5\%}}$ & \textbf{87.0}$^{\textcolor{mygreen}{\scriptsize +0.1\%}}$ \\
\midrule
\multicolumn{1}{l}{\multirow{5}{*}{MiniGPT-4-7B}} & Greedy    & 28.2 & 8.8 &\cellcolor{gray!15} 73.7 & 76.8 & 76.6 & 77.7 & 76.9 \\
& PAI               & 22.6 & 7.6 &\cellcolor{gray!15} 72.9 & 74.7 & 76.3 & 79.1 & 78.8 \\
& Devils            & 21.9 & 7.9 &\cellcolor{gray!15} 71.5 & 72.3 & 75.9 & 79.4 & 78.7 \\
& VISTA        & 18.0 & 4.3 &\cellcolor{mypink} 68.3 & 76.6 & 77.4 & ---  & ---  \\
& \textbf{HAVAE(Ours)} & \textbf{21.8}$^{\textcolor{mygreen}{\scriptsize -0.5\%}}$ & \textbf{6.9}$^{\textcolor{mygreen}{\scriptsize -9.2\%}}$ &\cellcolor{gray!15} 72.5 & \textbf{76.9}$^{\textcolor{mygreen}{\scriptsize +0.1\%}}$ & \textbf{77.6}$^{\textcolor{mygreen}{\scriptsize +0.3\%}}$ & \textbf{80.2}$^{\textcolor{mygreen}{\scriptsize +1.0\%}}$ & \textbf{80.2}$^{\textcolor{mygreen}{\scriptsize +1.8\%}}$ \\
\midrule
\multicolumn{1}{l}{\multirow{5}{*}{Shikra-7B}} & Greedy  & 56.8 & 14.8 &\cellcolor{gray!15} 75.4 & 81.0 & 81.7  & 76.4 & 78.3 \\
& PAI                 & 36.1 & 9.8  &\cellcolor{gray!15} 75.4 & 81.3 & 81.1 & 76.5 & 77.5 \\
& Devils              & 26.2 & 9.3  &\cellcolor{gray!15} 73.0 & 80.5 & 80.4 & 75.7 & 77.7 \\
& VISTA            & 32.8 & 9.8  &\cellcolor{gray!15} 73.4 & 81.3 & 81.9 & ---  & ---  \\
& \textbf{HAVAE(Ours)} & \textbf{15.8}$^{\textcolor{mygreen}{\scriptsize -39.7\%}}$ & \textbf{5.0}$^{\textcolor{mygreen}{\scriptsize -46.2\%}}$  &\cellcolor{gray!15} 71.8 & \textbf{81.6}$^{\textcolor{mygreen}{\scriptsize +0.4\%}}$ & \textbf{82.1}$^{\textcolor{mygreen}{\scriptsize +0.2\%}}$ & \textbf{76.7}$^{\textcolor{mygreen}{\scriptsize +0.3\%}}$ & \textbf{78.6}$^{\textcolor{mygreen}{\scriptsize +0.4\%}}$ \\
\midrule
\multicolumn{1}{l}{\multirow{3}{*}{LLaVA-1.5-13B}} & Greedy        & 41.6 & 11.1 & \cellcolor{gray!15} 79.3 & \textbf{82.6} & 84.5 & 85.4 & 83.2  \\
& Devils            & 29.0 & 8.6 & \cellcolor{gray!15} 79.9 & 71.4 & 77.2 & 87.8 & 86.4 \\
& \textbf{HAVAE(Ours)} & \textbf{21.8}$^{\textcolor{mygreen}{\scriptsize -24.8\%}}$ & \textbf{5.0}$^{\textcolor{mygreen}{\scriptsize -41.9\%}}$ & \cellcolor{gray!15} 79.8 & 82.5$^{\textcolor{myred}{\scriptsize -0.1\%}}$ & \textbf{84.7}$^{\textcolor{mygreen}{\scriptsize +0.2\%}}$ & \textbf{87.9}$^{\textcolor{mygreen}{\scriptsize +0.1\%}}$ & \textbf{86.6}$^{\textcolor{mygreen}{\scriptsize +0.2\%}}$  \\
\bottomrule
\end{tabular}
}
\caption{
Performance of \textbf{HAVAE(Ours)} against baselines. Best results are in \textbf{bold}. Pink cells mark potentially unreliable CHAIR scores. Superscripts show the \% change vs. the best baseline (excluding unreliable scores). 
}
\label{tab:main_results}
\end{table*}

\section{Hijacking-Aware Visual Attention Enhancement}
\label{sec:havae}

Leveraging the \textbf{NHAR} metric defined in \Cref{sec:correlation}, we propose \textbf{Hijacking-Aware Visual Attention Enhancement (HAVAE)}. 
This training-free framework operates in two stages to precisely identify and reinforce the attention heads critical for factual grounding, incurring zero additional computational overhead.

\noindent\textbf{Stage 1: Principled Head Selection.}
We first operationalize the NHAR metric to curate a set of critical heads. 
For each head $(\ell, h)$, we compute a global stability score, $\overline{\text{NHAR}}$, by averaging its performance across the real-object tokens ($\mathcal{O}_{\mathrm{real}}$) defined in \Cref{sec:case_setup}:
\begin{equation}
\label{eq:mean_nhar}
\overline{\text{NHAR}}^{(\ell,h)} = \frac{1}{|\mathcal{O}_{\mathrm{real}}|} \sum_{y_k \in \mathcal{O}_{\mathrm{real}}} \text{NHAR}^{(\ell,h)}(y_k).
\end{equation}
The top-$K$ heads ranked by $\overline{\text{NHAR}}$ constitute the target set $\mathcal{H}_{\text{target}}$, representing the model's primary channels for reliable visual processing.

\noindent\textbf{Stage 2: Collective Attention Reinforcement.}
During inference, we selectively intervene on $\mathcal{H}_{\text{target}}$. 
To amplify their focus without introducing external noise, we augment their attention weights using the layer-wise mean attention magnitude:
\begin{equation}
\label{eq:attention_enhancement}
\boldsymbol{A}_{k, i}^{(\ell, h)} \gets \boldsymbol{A}_{k, i}^{(\ell, h)} + \alpha \frac{1}{H} \sum_{h'=1}^{H} \left|\boldsymbol{A}_{k, i}^{(\ell, h')}\right|, \quad \forall v_i \in \mathcal{I}_v,
\end{equation}
where $\alpha$ controls the enhancement strength. 
By injecting this collective visual signal into the non-hijacked heads, HAVAE effectively counteracts Vocabulary Hijacking. 
The complete procedure is summarized in \Cref{alg:havae}.

\begin{table}[t]
\centering

\resizebox{\linewidth}{!}{%
\begin{tabular}{ll|cccc}
\toprule
\multicolumn{1}{l}{\multirow{2}{*}{\textbf{Metric}}} & 
\multicolumn{1}{l}{\multirow{2}{*}{\textbf{Method}}} & 
\textbf{LLaVA} & \textbf{MiniGPT-4} & \textbf{Shikra} & \textbf{LLaVA} \\
& & \textbf{1.5-7B} & \textbf{7B} & \textbf{7B} & \textbf{1.5-13B} \\
\midrule
\multirow{2}{*}{Perception} & Vanilla & 1472.5 & 731.9 & 962.0 & 1515.0 \\
 & \textbf{HAVAE (Ours)} & \textbf{1483.9} & \textbf{744.8} & \textbf{976.0} & \textbf{1528.9} \\
\midrule
\multirow{2}{*}{Cognition} & Vanilla & 322.5 & 173.0 & 250.4 & 290.6 \\
 & \textbf{HAVAE (Ours)} & \textbf{327.9} & \textbf{183.9} & \textbf{272.5} & \textbf{295.4} \\
\bottomrule
\end{tabular}
}
\caption{
MME Benchmark results. Comparison between vanilla models and \textbf{HAVAE (Ours)} on Perception and Cognition subsets.
}
\label{tab:mme_results}
\end{table}

\section{Experiments}
\label{sec:exp}
\subsection{Experimental Setup}
\label{sec:exp_setup}

\noindent\textbf{Models.} We evaluate our method on LLaVA-1.5 (7B, 13B)~\citep{liu2023llava}, Shikra (7B)~\citep{shikra}, MiniGPT-4 (7B)~\citep{minigpt4}, and Qwen2-VL (7B)~\citep{wang2024qwen2} to assess its generalizability and scalability.

\noindent\textbf{Evaluation Benchmarks.} Following prior work~\citep{liu2024paying, jiang2025devils}, we evaluate our method on several hallucination benchmarks. 
We use \textbf{CHAIR}~\citep{rohrbach2018object} for captioning evaluation and \textbf{POPE}~\citep{pope} for query-based object probing, along with its conversational variant, \textbf{POPE-Chat}. 
To assess out-of-domain performance and fine-grained errors, we also employ \textbf{AMBER}~\citep{wang2023llm}. 
Finally, to investigate the impact on general capabilities, we conduct evaluations on \textbf{MME}~\citep{bench:mme}. 
Further details are provided in \Cref{app:benchmark}.

\begin{table}[t]
\centering
\resizebox{\linewidth}{!}{%
\begin{tabular}{ll|ccc|cc|c}
\toprule
\multicolumn{1}{l}{\multirow{2}{*}{\textbf{Model}}} &
\multicolumn{1}{l}{\multirow{2}{*}{\textbf{Method}}} &
\multicolumn{3}{c}{\textbf{CHAIR}} &
\multicolumn{2}{c}{\textbf{POPE}} &
\multicolumn{1}{c}{\textbf{MME}} \\
\cmidrule(lr){3-5} \cmidrule(lr){6-7} \cmidrule(lr){8-8}
& & \textbf{CHAIR$_s$ $\downarrow$} & \textbf{CHAIR$_i$ $\downarrow$} & \textbf{F1 $\uparrow$} & \textbf{Acc. $\uparrow$} & \textbf{F1 $\uparrow$} & \textbf{All $\uparrow$} \\
\midrule
\multirow{2}{*}{Qwen2-VL} & Vanilla & 27.6 & 8.8 & 79.2 & 89.1 & 88.1 & 2268.4 \\
 & \textbf{HAVAE} & \textbf{22.8} & \textbf{6.2} & \textbf{79.4} & \textbf{89.3} & \textbf{88.9} & \textbf{2290.2} \\
\bottomrule
\end{tabular}
}
\caption{
Performance comparison on Qwen2-VL across CHAIR, POPE, and MME benchmarks.
}
\label{tab:qwen2vl_results}
\end{table}

\begin{figure*}[t]
    \centering
    \includegraphics[width=\linewidth]{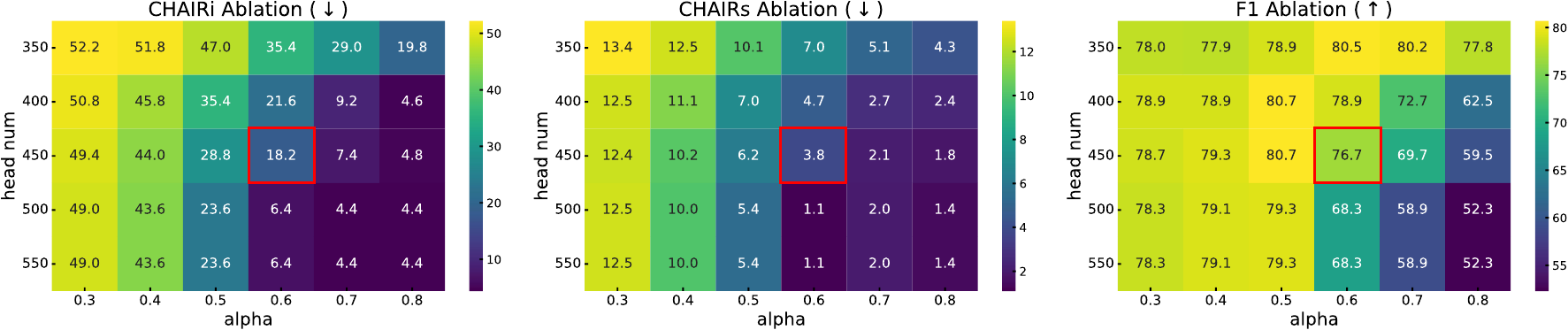}
    \caption{Ablation on hyperparameters $\alpha$ and $K$ for LLaVA-1.5 7B. Red boxes highlight the parameter combinations we used.}
    \vspace{-3mm}
\label{fig:llava_abaltion}
\end{figure*}
\begin{table*}[htbp!]
\centering
\resizebox{\textwidth}{!}{%
\begin{tabular}{ll|ccc|cc|cc|cc}
\toprule
\multicolumn{1}{l}{\multirow{2}{*}{\textbf{Model}}}
&\multicolumn{1}{l}{\multirow{2}{*}{\textbf{Selecting Strategy}}}
& \multicolumn{3}{c}{\textbf{CHAIR}} 
& \multicolumn{2}{c}{\textbf{POPE}} 
& \multicolumn{2}{c}{\textbf{POPE Chat}} 
& \multicolumn{2}{c}{\textbf{MME}} \\ 
\cmidrule(lr){3-5} \cmidrule(lr){6-7} \cmidrule(lr){8-9} \cmidrule(lr){10-11}
& & \textbf{CHAIR$_s$ $\downarrow$} & \textbf{CHAIR$_i$ $\downarrow$}& \textbf{F1 $\uparrow$} & \textbf{Acc. $\uparrow$} & \textbf{F1 $\uparrow$} & \textbf{Acc. $\uparrow$} & \textbf{F1 $\uparrow$} & \textbf{Per. $\uparrow$} & \textbf{Cog. $\uparrow$} \\
\midrule
\multicolumn{1}{l}{\multirow{2}{*}{LLaVA-1.5-7B}} & Max Attention         & 7.8 & 4.4 &\cellcolor{mypink}65.8 & 85.9 & 85.6 & 86.0 & 85.5 & 1399.0 & 277.0 \\
& \textbf{HAVAE(Ours)} & 18.2 & 3.8 &\cellcolor{gray!15} 76.7 & \textbf{86.2} & \textbf{86.3} & \textbf{88.0} & \textbf{87.0} & \textbf{1483.9} & \textbf{327.9} \\
\bottomrule
\end{tabular}
}
\caption{Comparison of head selection strategies on LLaVA-1.5 7B.}
\label{tab:head}
\end{table*}
\noindent\textbf{Baselines.} 
We evaluate HAVAE against representative training-free hallucination mitigation strategies. 
Our primary comparisons focus on attention-based interventions, specifically \textbf{PAI}~\citep{liu2024paying} and \textbf{Devils}~\citep{jiang2025devils}, which are most relevant to our approach. 
Additionally, we evaluate against \textbf{VISTA}~\citep{lihidden}, a leading method for activation steering.

\noindent\textbf{Implementation Details.} 
We set the head count $K=300$ for Qwen2-VL and $K=450$ for all other models. 
The enhancement strength $\alpha$ defaults to 0.1, but for long-context tasks (Captioning, POPE-Chat), it is increased to 0.6 (LLaVA-1.5-7B, MiniGPT-4, Qwen2-VL) or 0.7 (LLaVA-1.5-13B, Shikra). 
We adopt greedy decoding as the default strategy for our main experiments. Comprehensive results using alternative decoding methods are provided in \Cref{app:decoding}.

\subsection{Main Results}
\label{sec:main_results}

\noindent\textbf{Superior Performance and Generalization.}
As detailed in \Cref{tab:main_results}, HAVAE demonstrates clear superiority across all evaluated tasks. 
In standard settings, it excels in both long-form (CHAIR) and short-form (POPE) generation, reducing the CHAIR\textsubscript{i} score by 38.7\% on LLaVA-1.5-7B and 46.2\% on Shikra-7B relative to the strongest reliable baseline. 
Crucially, this effectiveness extends to scenarios beyond the calibration domain. 
When applied to the AMBER benchmark—which contains images distinct from the COCO source used for head selection—HAVAE maintains significant gains (\Cref{tab:amber_combined}, see \Cref{app:amber}). 
Furthermore, even in the absence of ground-truth annotations for calibration (analyzed in \Cref{sec:ablation_no_gt}), our method sustains high performance.
These results confirm that HAVAE captures intrinsic, model-specific anti-hallucination mechanisms rather than overfitting to specific datasets or supervision signals.

\noindent\textbf{Preserving General Capabilities.}
To verify that HAVAE preserves fundamental capabilities, we evaluate on the comprehensive MME benchmark. 
As shown in \Cref{tab:mme_results}, HAVAE consistently surpasses vanilla baselines across all tested architectures. 
This suggests that by suppressing the attention noise of Inert Tokens, our method mitigates hallucinations while maintaining—and often enhancing—general visual perception and reasoning.

\noindent\textbf{Scalability and Architectural Generalization.}
To assess the robustness of our approach, we extend our evaluation to larger scales and advanced architectures. 
As shown in \Cref{tab:main_results}, HAVAE yields significant improvements on LLaVA-1.5 13B, confirming its scalability. 
Furthermore, validation on the advanced Qwen2-VL 7B (\Cref{tab:qwen2vl_results}) demonstrates consistent gains over baselines. 
These combined results confirm that HAVAE effectively generalizes across diverse model sizes and structural designs.


\begin{figure*}[t]
    \centering
    \includegraphics[width=0.95\linewidth]{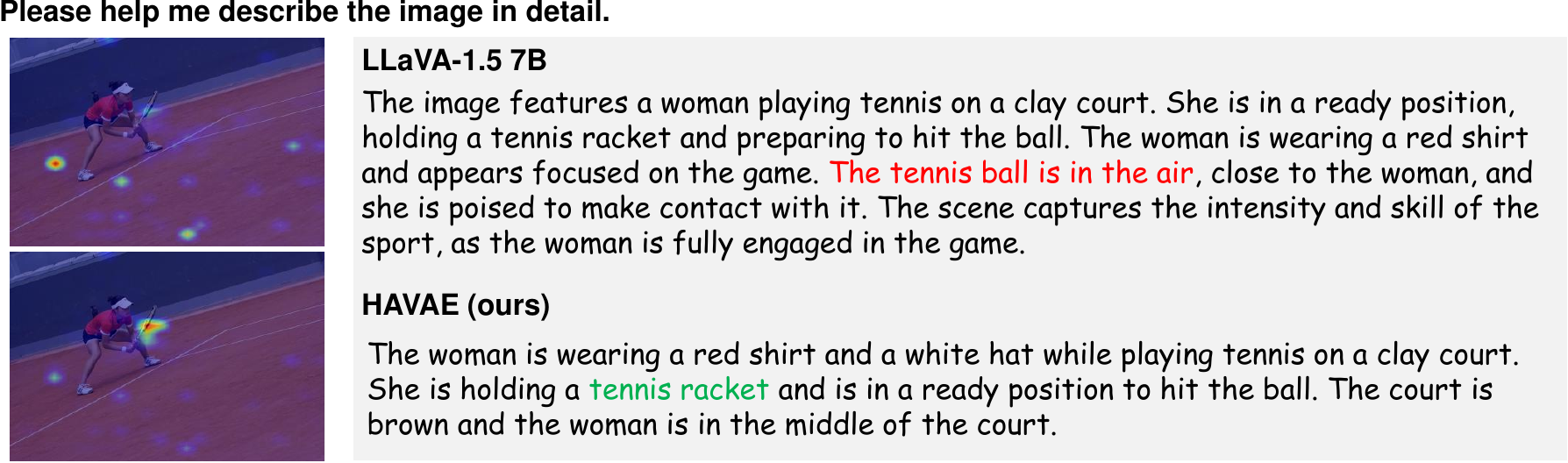}
    \caption{A case study demonstrates HAVAE correcting a baseline hallucination by redirecting the model's focus from irrelevant Inert tokens (as shown in the heatmap) back to the salient visual target.}
    \label{fig:case}
\end{figure*}

\subsection{Ablation Studies and Analysis}

\noindent\textbf{Impact of Hyperparameters $\alpha$ and $K$.}
We investigate $\alpha \in [0.3, 0.8]$ and $K \in [350, 550]$, observing that precise tuning is particularly critical in long-context scenarios.
Results on LLaVA-1.5 7B (\Cref{fig:llava_abaltion}) reveal a trade-off: aggressive settings effectively reduce hallucinations (lower CHAIR) but degrade generation quality (lower F1).
Consequently, we select parameters to balance suppression with fluency. 
Additional ablations are provided in \Cref{app:albation}.

\begin{table}[t]
    \centering
    \resizebox{0.9\linewidth}{!}{%
    \begin{tabular}{c ccc ccc}
        \toprule
        & \multicolumn{3}{c}{\textbf{CHAIR}} & \multicolumn{2}{c}{\textbf{POPE}} \\
        \cmidrule(lr){2-4} \cmidrule(lr){5-6}
        \textbf{Num} & \textbf{CHAIR$_s$} & \textbf{CHAIR$_i$} & \textbf{F1} & \textbf{Acc} & \textbf{F1} \\
        \midrule
        10   & 18.8 & 3.7 & \cellcolor{gray!15}76.5 & 86.1 & 86.2 \\
        100  & 18.8 & 3.7 & \cellcolor{gray!15}76.5 & 85.9 & 86.1 \\
        300  & 18.6 & 4.8 & \cellcolor{gray!15}76.9 & 85.9 & 86.1 \\
        500  & 18.2 & 3.7 & \cellcolor{gray!15}76.7 & 86.1 & 86.2 \\
        1000 & 18.2 & 3.7 & \cellcolor{gray!15}76.7 & 86.1 & 86.2 \\
        \bottomrule
    \end{tabular}
    }
    \caption{Ablation study on the sample size for LLaVA-1.5 7B.}
    \label{tab:sample}
\end{table}

\noindent\textbf{Superiority of NHAR for Head Selection.}
We demonstrate NHAR's superiority by ablating its core component: hijacking-awareness. 
As shown in \Cref{tab:head}, selecting heads based only on total visual attention ($\mathcal{I}_{\text{inert}} = \emptyset$) \textbf{significantly lowers F1 scores} across all benchmarks on LLaVA-1.5 7B. 
This confirms high raw visual attention is a flawed proxy for effective grounding, easily captured by redundant Inert Tokens. 
NHAR's crucial advantage lies in filtering this noise, isolating heads that truly focus on salient visual content.

\noindent\textbf{Ablation on Sample Size.}
We performed an ablation on the number of samples used for our analysis in \Cref{sec:case_setup}. As shown in \Cref{tab:sample}, key performance metrics across all benchmarks stabilize once the sample size reaches 500. Therefore, we adopt a sample size of 500 for all experiments, as this provides a robust estimate of performance without incurring unnecessary computational costs.

\noindent\textbf{Effect of Penalizing Attention to Inert Tokens.}
Positing that \textbf{Inert Tokens} are redundant, we hypothesized that explicitly penalizing their attention might improve performance. 
We introduced a penalty factor $\beta$, where $\beta=0$ corresponds to the standard HAVAE configuration. 
As $\beta$ increases, the suppression of attention on Inert Tokens progressively intensifies. 
However, results in \Cref{tab:beta} indicate that this strategy offers no benefit and is in fact detrimental, causing notable performance degradation on CHAIR. 
This observation, consistent consistent with prior work~\citep{guattention}, confirms that suppressing attention to these tokens impairs generation, thereby validating our design choice to employ positive attention reinforcement rather than negative suppression.

\noindent\textbf{Origins of Vocabulary Hijacking.}
We hypothesize that this vocabulary hijacking is an artifact deeply rooted in the initial vision-language alignment process. When we analyzed the base LLaVA-1.5 model (which only underwent multimodal pre-training without subsequent instruction tuning), we observed that the vocabulary hijacking phenomenon was significantly more severe. In this early stage, almost all attention weights were monopolized by these Inert tokens. This suggests that during pre-training, the model may learn a shortcut to dump unaligned or redundant visual features into specific vocabulary anchors.

\begin{table}[t]
    \centering
    \resizebox{0.86\linewidth}{!}{%
    \begin{tabular}{c ccc ccc}
        \toprule
        & \multicolumn{3}{c}{\textbf{CHAIR}} & \multicolumn{2}{c}{\textbf{POPE}} \\
        \cmidrule(lr){2-4} \cmidrule(lr){5-6}
        $\beta$ & \textbf{CHAIR$_s$} & \textbf{CHAIR$_i$} & \textbf{F1} & \textbf{Acc} & \textbf{F1} \\
        \midrule
        0.0 & 18.2 & 3.7 & \cellcolor{gray!15}76.7 & 86.1 & 86.2 \\
        0.1 & 20.2 & 4.5 & \cellcolor{gray!15}76.8 & 86.1 & 86.2 \\
        0.3 & 19.0 & 4.1 & \cellcolor{gray!15}77.0 & 86.2 & 86.2 \\
        0.6 & 19.8 & 4.7 & \cellcolor{gray!15}76.8 & 86.1 & 86.2 \\
        0.9 & 19.2 & 4.7 & \cellcolor{gray!15}76.9 & 86.2 & 86.3 \\
        \bottomrule
    \end{tabular}
    }
    \caption{Ablation on the sink token penalty factor $\beta$ for LLaVA-1.5 7B.}
    \label{tab:beta}
\end{table}

\subsection{Hyperparameter Sensitivity Analysis}
\label{app:hyperparameter_sensitivity}

To quantitatively evaluate the stability of our Hijacking Anchor-Based Identification (HABI) metric and the overall HAVAE method, we conduct ablation studies by varying the identification thresholds $\tau_r$ and $\tau_s$. As shown in \Cref{tab:sensitivity_tau_r} and \Cref{tab:sensitivity_tau_s}, we scale the thresholds from $0.8\times$ to $1.2\times$ of our empirically established optimal values ($\tau^*$). 

The experimental results clearly indicate that the performance remains highly stable under these fluctuations. This analysis confirms that the proposed method is robust and not overly sensitive to minor variations in hyperparameter selection.

Due to space limitations, additional experimental results and analyses are presented in \Cref{app:analysis}.

\begin{table}[t]
    \centering
    \resizebox{0.9\linewidth}{!}{%
    \begin{tabular}{c ccc cc}
        \toprule
        & \multicolumn{3}{c}{\textbf{CHAIR}} & \multicolumn{2}{c}{\textbf{POPE}} \\
        \cmidrule(lr){2-4} \cmidrule(lr){5-6}
        \textbf{$\tau_r / \tau_r^*$} & \textbf{CHAIR$_s$} & \textbf{CHAIR$_i$} & \textbf{F1} & \textbf{Acc} & \textbf{F1} \\
        \midrule
        0.8 & 18.5 & 4.3 & \cellcolor{gray!15}76.9 & 86.3 & 86.2 \\
        0.9 & 18.0 & 4.0 & \cellcolor{gray!15}76.9 & 86.1 & 86.4 \\
        \textbf{1.0 (Ours)} & 18.2 & 3.8 & \cellcolor{gray!15}76.7 & 86.2 & 86.3 \\
        1.1 & 17.9 & 3.9 & \cellcolor{gray!15}76.8 & 86.1 & 86.3 \\
        1.2 & 18.3 & 4.1 & \cellcolor{gray!15}76.6 & 86.2 & 86.3 \\
        \bottomrule
    \end{tabular}
    }
    \caption{Sensitivity Analysis of $\tau_r$. HAVAE exhibits strong robustness to variations in the $\tau_r$ threshold.}
    \label{tab:sensitivity_tau_r}
\end{table}

\subsection{Case Study}
\label{sec:case}

A case study in \Cref{fig:case} (d) illustrates our method's effectiveness. 
While the baseline hallucinates that ``The tennis ball is in the air'', our method provides a factually accurate description. 
The attention visualizations reveal the mechanism behind this correction: the baseline, when generating \textbf{``ball''}, suffers from severe \textbf{Vocabulary Hijacking}, where its attention is captured by \textbf{Inert Tokens} and scattered onto irrelevant background. 
In contrast, the heads selected by HAVAE demonstrate a precise focus on the target object when generating \textbf{``racket''}, thereby preventing the hallucination. 
Additional case studies are presented in \Cref{app:case}.

\section{Conclusion}

This paper provides the first mechanistic analysis of the \textbf{Vocabulary Hijacking} phenomenon in LVLMs. 
We reveal a core property where specific visual tokens, termed \textbf{Inert Tokens}, hijack the model's attention and consistently decode to a fixed set of \textbf{Hijacking Anchors}. 
This discovery forms the foundation for our \textbf{Hijacking Anchor-Based Identification (HABI)} method, which reliably localizes these \textbf{Inert Tokens}. 
Building on this, we propose \textbf{Hijacking-Aware Visual Attention Enhancement (HAVAE)}, a training-free intervention that leverages our \textbf{Non-Hijacked Visual Attention Ratio (NHAR)} metric to identify hallucination-critical heads and selectively strengthen their focus on salient visual content. 
Our experimental results demonstrate that HAVAE significantly reduces hallucination with \textbf{zero computational overhead}, while preserving the model's general capabilities. 
We believe our work contributes to a deeper, mechanistic understanding of attention anomalies and their causal link to hallucination, offering a new direction for improving the reliability of LVLMs.

\begin{table}[t]
    \centering
    \resizebox{0.9\linewidth}{!}{%
    \begin{tabular}{c ccc cc}
        \toprule
        & \multicolumn{3}{c}{\textbf{CHAIR}} & \multicolumn{2}{c}{\textbf{POPE}} \\
        \cmidrule(lr){2-4} \cmidrule(lr){5-6}
        \textbf{$\tau_s / \tau_s^*$} & \textbf{CHAIR$_s$} & \textbf{CHAIR$_i$} & \textbf{F1} & \textbf{Acc} & \textbf{F1} \\
        \midrule
        0.8 & 18.6 & 4.5 & \cellcolor{gray!15}76.6 & 86.2 & 86.3 \\
        0.9 & 17.8 & 3.9 & \cellcolor{gray!15}76.8 & 86.3 & 86.2 \\
        \textbf{1.0 (Ours)} & 18.2 & 3.8 & \cellcolor{gray!15}76.7 & 86.2 & 86.3 \\
        1.1 & 18.1 & 4.1 & \cellcolor{gray!15}77.0 & 86.3 & 86.2 \\
        1.2 & 18.4 & 4.2 & \cellcolor{gray!15}76.9 & 86.1 & 86.3 \\
        \bottomrule
    \end{tabular}
    }
    \caption{Sensitivity Analysis of $\tau_s$. HAVAE exhibits strong robustness to variations in the $\tau_s$ threshold.}
    \label{tab:sensitivity_tau_s}
\end{table}

\section*{Acknowledgements}
This work was supported in part by National Natural Science Foundation of China (62476070), Shenzhen Science and Technology Program \seqsplit{(JCYJ20241202123503005, \, GXWD20231128103232001, \,ZDSYS20230626091203008,\, KQTD20240729102154066)}, Department of Science and Technology of Guangdong (2024A1515011540), National Key R\&D Program of China (SQ2024YFE0200592) and Suzhou Science and Technology Program (SYG2025072).

\section*{Limitations}

First, while this work uncovers the phenomenon of Vocabulary Hijacking, its mechanistic origin remains to be fully elucidated. 
Specifically, the precise training dynamics that precipitate this behavior remain unclear. 
Moreover, understanding why a distinct subset of semantically meaningless tokens crystallize into Hijacking Anchors is a complex question we leave for future investigation. 
Second, constrained by computational resources, our experimental validation is currently limited to models with sizes up to 13B parameters. 
Extending this analysis to larger-scale foundational models represents a vital direction for follow-up research. 
Finally, exploring training-based strategies to mitigate Vocabulary Hijacking constitutes another critical avenue for future work.

\section*{Ethical Considerations}

Our method, HAVAE, is a training-free intervention validated on standard publicly available benchmarks (e.g., COCO, POPE). 
We explicitly state that the efficacy of our findings may be influenced by the distribution of these evaluation datasets. 
Consequently, the performance of our method on highly specialized domains or privacy-sensitive data remains unverified. 
Thus, potential risks exist when deploying HAVAE in high-stakes applications (e.g., medical or legal consultation). 
We strongly advise users to exercise caution and maintain human oversight to verify the factual correctness of the model's responses, as the mitigation of hallucination does not guarantee the complete elimination of errors.

\bibliography{custom}

\begin{thebibliography}{35}
\providecommand{\natexlab}[1]{#1}

\bibitem[{Bai et~al.(2024)Bai, Wang, Xiao, He, Han, Zhang, and
  Shou}]{bai2024hallucination}
Zechen Bai, Pichao Wang, Tianjun Xiao, Tong He, Zongbo Han, Zheng Zhang, and
  Mike~Zheng Shou. 2024.
\newblock Hallucination of multimodal large language models: A survey.
\newblock \emph{arXiv preprint arXiv:2404.18930}.

\bibitem[{Chen et~al.(2023)Chen, Zhang, Zeng, Zhang, Zhu, and Zhao}]{shikra}
Keqin Chen, Zhao Zhang, Weili Zeng, Richong Zhang, Feng Zhu, and Rui Zhao.
  2023.
\newblock Shikra: Unleashing multimodal llm's referential dialogue magic.
\newblock \emph{arXiv preprint arXiv:2306.15195}.

\bibitem[{Chen et~al.(2024)Chen, Zhao, Luo, Yao, Li, and Zhou}]{chen2024halc}
Zhaorun Chen, Zhuokai Zhao, Hongyin Luo, Huaxiu Yao, Bo~Li, and Jiawei Zhou.
  2024.
\newblock Halc: Object hallucination reduction via adaptive focal-contrast
  decoding.
\newblock In \emph{Proceedings of the International Conference on Machine
  Learning (ICML)}.

\bibitem[{Deiseroth et~al.(2023)Deiseroth, Deb, Weinbach, Brack, Schramowski,
  and Kersting}]{Deiseroth2023atman}
Bj\"{o}rn Deiseroth, Mayukh Deb, Samuel Weinbach, Manuel Brack, Patrick
  Schramowski, and Kristian Kersting. 2023.
\newblock Atman: Understanding transformer predictions through memory efficient
  attention manipulation.
\newblock In \emph{Advances in Neural Information Processing Systems
  (NeurIPS)}, volume~36, pages 63437--63460. Curran Associates, Inc.

\bibitem[{Du et~al.(2025)Du, Fang, Li, Li, Jiang, Yu, Guo, Chen, Goh, Tang, He,
  Liu, and Zhang}]{du2025nps}
Guodong Du, Zitao Fang, Jing Li, Junlin Li, Runhua Jiang, Shuyang Yu, Yifei
  Guo, Yangneng Chen, Sim~Kuan Goh, Ho-Kin Tang, Daojing He, Honghai Liu, and
  Min Zhang. 2025.
\newblock \href {https://doi.org/10.18653/v1/2025.acl-long.1570} {Neural
  parameter search for slimmer fine-tuned models and better transfer}.
\newblock In \emph{Proceedings of the 63rd Annual Meeting of the Association
  for Computational Linguistics (Volume 1: Long Papers)}.

\bibitem[{Du et~al.(2024)Du, Lee, Li, Jiang, Guo, Yu, Liu, Goh, Tang, He, and
  Zhang}]{du2024pcb}
Guodong Du, Junlin Lee, Jing Li, Runhua Jiang, Yifei Guo, Shuyang Yu, Hanting
  Liu, Sim~Kuan Goh, Ho-Kin Tang, Daojing He, and Min Zhang. 2024.
\newblock Parameter competition balancing for model merging.
\newblock In \emph{The Thirty-eighth Annual Conference on Neural Information
  Processing Systems (NeurIPS)}.

\bibitem[{Du et~al.(2026)Du, Li, Zhou, Li, Shi, Lin, Tang, Li, Liu, Wang,
  Zhang, and Li}]{du2025graftllm}
Guodong Du, Zhuo Li, Xuanning Zhou, Junlin Li, Zesheng Shi, Wanyu Lin, Ho-Kin
  Tang, Xiucheng Li, Fangming Liu, Wenya Wang, Min Zhang, and Jing Li. 2026.
\newblock Knowledge fusion of large language models via modular skillpacks.
\newblock In \emph{Proceedings of the International Conference on Learning
  Representations (ICLR)}.

\bibitem[{Elhage et~al.(2021)Elhage, Nanda, Olsson, Henighan, Joseph, Mann,
  Askell, Bai, Chen, Conerly, DasSarma, Drain, Ganguli, Hatfield-Dodds,
  Hernandez, Jones, Kernion, Lovitt, Ndousse, Amodei, Brown, Clark, Kaplan,
  McCandlish, and Olah}]{elhage2021mathematical}
Nelson Elhage, Neel Nanda, Catherine Olsson, Tom Henighan, Nicholas Joseph, Ben
  Mann, Amanda Askell, Yuntao Bai, Anna Chen, Tom Conerly, Nova DasSarma, Dawn
  Drain, Deep Ganguli, Zac Hatfield-Dodds, Danny Hernandez, Andy Jones, Jackson
  Kernion, Liane Lovitt, Kamal Ndousse, and 6 others. 2021.
\newblock A mathematical framework for transformer circuits.
\newblock \emph{Transformer Circuits Thread}, 1(1):12.

\bibitem[{Fu et~al.(2023)Fu, Chen, Shen, Qin, Zhang, Lin, Yang, Zheng, Li, Sun
  et~al.}]{bench:mme}
Chaoyou Fu, Peixian Chen, Yunhang Shen, Yulei Qin, Mengdan Zhang, Xu~Lin,
  Jinrui Yang, Xiawu Zheng, Ke~Li, Xing Sun, and 1 others. 2023.
\newblock Mme: A comprehensive evaluation benchmark for multimodal large
  language models.
\newblock \emph{arXiv preprint arXiv:2306.13394}.

\bibitem[{Ge et~al.(2024)Ge, Zhang, Liu, Zhang, Han, and Gao}]{ge2024model}
Suyu Ge, Yunan Zhang, Liyuan Liu, Minjia Zhang, Jiawei Han, and Jianfeng Gao.
  2024.
\newblock Model tells you what to discard: Adaptive {KV} cache compression for
  {LLM}s.
\newblock In \emph{The Twelfth International Conference on Learning
  Representations (ICLR)}.

\bibitem[{Gu et~al.(2025)Gu, Pang, Du, Liu, Zhang, Du, Wang, and
  Lin}]{guattention}
Xiangming Gu, Tianyu Pang, Chao Du, Qian Liu, Fengzhuo Zhang, Cunxiao Du,
  Ye~Wang, and Min Lin. 2025.
\newblock When attention sink emerges in language models: An empirical view.
\newblock In \emph{The Thirteenth International Conference on Learning
  Representations (ICLR)}.

\bibitem[{Huang et~al.(2024)Huang, Dong, Zhang, Wang, He, Wang, Lin, Zhang, and
  Yu}]{huang2023opera}
Qidong Huang, Xiaoyi Dong, Pan Zhang, Bin Wang, Conghui He, Jiaqi Wang, Dahua
  Lin, Weiming Zhang, and Nenghai Yu. 2024.
\newblock {OPERA:} alleviating hallucination in multi-modal large language
  models via over-trust penalty and retrospection-allocation.
\newblock In \emph{Proceedings of the IEEE/CVF Conference on Computer Vision
  and Pattern Recognition (CVPR)}, pages 13418--13427.

\bibitem[{Jiang et~al.(2025)Jiang, Chen, Zhu, Luo, Shen, and
  Yang}]{jiang2025devils}
Zhangqi Jiang, Junkai Chen, Beier Zhu, Tingjin Luo, Yankun Shen, and Xu~Yang.
  2025.
\newblock Devils in middle layers of large vision-language models:
  Interpreting, detecting and mitigating object hallucinations via attention
  lens.
\newblock In \emph{Proceedings of the IEEE/CVF Conference on Computer Vision
  and Pattern Recognition (CVPR)}, pages 25004--25014.

\bibitem[{Kang et~al.(2025)Kang, Kim, Kim, and Hwang}]{visattnsink}
Seil Kang, Jinyeong Kim, Junhyeok Kim, and Seong~Jae Hwang. 2025.
\newblock See what you are told: Visual attention sink in large multimodal
  models.
\newblock In \emph{The Thirteenth International Conference on Learning
  Representations, {ICLR} 2025, Singapore, April 24-28, 2025}.

\bibitem[{Leng et~al.(2024)Leng, Zhang, Chen, Li, Lu, Miao, and Bing}]{VCD}
Sicong Leng, Hang Zhang, Guanzheng Chen, Xin Li, Shijian Lu, Chunyan Miao, and
  Lidong Bing. 2024.
\newblock Mitigating object hallucinations in large vision-language models
  through visual contrastive decoding.
\newblock In \emph{Proceedings of the IEEE/CVF Conference on Computer Vision
  and Pattern Recognition (CVPR)}, pages 13872--13882.

\bibitem[{Li et~al.(2023)Li, Du, Zhou, Wang, Zhao, and Wen}]{pope}
Yifan Li, Yifan Du, Kun Zhou, Jinpeng Wang, Wayne~Xin Zhao, and Ji-Rong Wen.
  2023.
\newblock Evaluating object hallucination in large vision-language models.
\newblock \emph{arXiv preprint arXiv:2305.10355}.

\bibitem[{Li et~al.(2025)Li, Shi, Gao, Liu, Wang, Chen, Liu, Zhao, Wang, and
  Metaxas}]{lihidden}
Zhuowei Li, Haizhou Shi, Yunhe Gao, Di~Liu, Zhenting Wang, Yuxiao Chen, Ting
  Liu, Long Zhao, Hao Wang, and Dimitris~N Metaxas. 2025.
\newblock The hidden life of tokens: Reducing hallucination of large
  vision-language models via visual information steering.
\newblock In \emph{Forty-second International Conference on Machine Learning
  (ICLR)}.

\bibitem[{Lin et~al.(2014)Lin, Maire, Belongie, Hays, Perona, Ramanan,
  Doll{\'a}r, and Zitnick}]{lin2014mscoco}
Tsung-Yi Lin, Michael Maire, Serge Belongie, James Hays, Pietro Perona, Deva
  Ramanan, Piotr Doll{\'a}r, and C~Lawrence Zitnick. 2014.
\newblock Microsoft coco: Common objects in context.
\newblock In \emph{Computer Vision--ECCV 2014: 13th European Conference,
  Zurich, Switzerland, September 6-12, 2014}.

\bibitem[{Liu et~al.(2023)Liu, Li, Wu, and Lee}]{liu2023llava}
Haotian Liu, Chunyuan Li, Qingyang Wu, and Yong~Jae Lee. 2023.
\newblock Visual instruction tuning.
\newblock In \emph{Proceedings of the Advances in Neural Information Processing
  Systems (NeurIPS)}.

\bibitem[{Liu et~al.(2024)Liu, Zheng, and Chen}]{liu2024paying}
Shi Liu, Kecheng Zheng, and Wei Chen. 2024.
\newblock Paying more attention to image: A training-free method for
  alleviating hallucination in lvlms.
\newblock In \emph{Proceedings of the European Conference on Computer Vision
  (ECCV)}, pages 125--140.

\bibitem[{nostalgebraist(2020)}]{logitLens}
nostalgebraist. 2020.
\newblock Interpreting gpt: The logit lens.
\newblock
  \url{https://www.lesswrong.com/posts/AcKRB8wDpdaN6v6ru/interpreting-gpt-the-logit-lens}.

\bibitem[{Otsu(1979)}]{otsu1979threshold}
Nobuyuki Otsu. 1979.
\newblock A threshold selection method from gray-level histograms.
\newblock \emph{IEEE Transactions on Systems, Man, and Cybernetics},
  9(1):62--66.

\bibitem[{Rohrbach et~al.(2018)Rohrbach, Hendricks, Burns, Darrell, and
  Saenko}]{rohrbach2018object}
Anna Rohrbach, Lisa~Anne Hendricks, Kaylee Burns, Trevor Darrell, and Kate
  Saenko. 2018.
\newblock Object hallucination in image captioning.
\newblock In \emph{Proceedings of the Conference on Empirical Methods in
  Natural Language Processing (EMNLP)}, pages 4035--4045.

\bibitem[{Sun et~al.(2024)Sun, Shen, Cao, Liu, Li, Shen, Gan, Gui, Wang, Yang,
  Keutzer, and Darrell}]{llavarlhf}
Zhiqing Sun, Sheng Shen, Shengcao Cao, Haotian Liu, Chunyuan Li, Yikang Shen,
  Chuang Gan, Liang{-}Yan Gui, Yu{-}Xiong Wang, Yiming Yang, Kurt Keutzer, and
  Trevor Darrell. 2024.
\newblock Aligning large multimodal models with factually augmented {RLHF}.
\newblock In \emph{Proceedings of the Annual Meeting of the Association for
  Computational Linguistics (ACL)}, pages 13088--13110.

\bibitem[{Wang et~al.(2023)Wang, Wang, Xu, Zhang, Gu, Jia, Yan, Zhang, and
  Sang}]{wang2023llm}
Junyang Wang, Yuhang Wang, Guohai Xu, Jing Zhang, Yukai Gu, Haitao Jia, Ming
  Yan, Ji~Zhang, and Jitao Sang. 2023.
\newblock An llm-free multi-dimensional benchmark for mllms hallucination
  evaluation.
\newblock \emph{arXiv preprint arXiv:2311.07397}.

\bibitem[{Wang et~al.(2024)Wang, Bai, Tan, Wang, Fan, Bai, Chen, Liu, Wang, Ge
  et~al.}]{wang2024qwen2}
Peng Wang, Shuai Bai, Sinan Tan, Shijie Wang, Zhihao Fan, Jinze Bai, Keqin
  Chen, Xuejing Liu, Jialin Wang, Wenbin Ge, and 1 others. 2024.
\newblock Qwen2-vl: Enhancing vision-language model's perception of the world
  at any resolution.
\newblock \emph{arXiv preprint arXiv:2409.12191}.

\bibitem[{Woo et~al.(2025)Woo, Kim, Jang, Choi, and Kim}]{AVISC}
Sangmin Woo, Donguk Kim, Jaehyuk Jang, Yubin Choi, and Changick Kim. 2025.
\newblock Don’t miss the forest for the trees: Attentional vision calibration
  for large vision language models.
\newblock In \emph{Findings of the Association for Computational Linguistics:
  ACL 2025}.

\bibitem[{Xiao et~al.(2023)Xiao, Tian, Chen, Han, and
  Lewis}]{xiao23streamingllm}
Guangxuan Xiao, Yuandong Tian, Beidi Chen, Song Han, and Mike Lewis. 2023.
\newblock Efficient streaming language models with attention sinks.
\newblock \emph{arXiv preprint arXiv:2309.17453}.

\bibitem[{Yang et~al.(2025)Yang, Li, Cao, and Xu}]{yang2025mitigating}
Tianyun Yang, Ziniu Li, Juan Cao, and Chang Xu. 2025.
\newblock Mitigating hallucination in large vision-language models via modular
  attribution and intervention.
\newblock In \emph{Adaptive Foundation Models: Evolving AI for Personalized and
  Efficient Learning}.

\bibitem[{Yu et~al.(2024{\natexlab{a}})Yu, Yao, Zhang, He, Han, Cui, Hu, Liu,
  Zheng, Sun, and Chua}]{2023rlhf-v}
Tianyu Yu, Yuan Yao, Haoye Zhang, Taiwen He, Yifeng Han, Ganqu Cui, Jinyi Hu,
  Zhiyuan Liu, Hai{-}Tao Zheng, Maosong Sun, and Tat{-}Seng Chua.
  2024{\natexlab{a}}.
\newblock {RLHF-V:} towards trustworthy mllms via behavior alignment from
  fine-grained correctional human feedback.
\newblock In \emph{Proceedings of the IEEE/CVF Conference on Computer Vision
  and Pattern Recognition (CVPR)}, pages 13807--13816.

\bibitem[{Yu et~al.(2024{\natexlab{b}})Yu, Zhang, Yao, Dang, Chen, Lu, Cui, He,
  Liu, Chua, and Sun}]{yu2024rlaifv}
Tianyu Yu, Haoye Zhang, Yuan Yao, Yunkai Dang, Da~Chen, Xiaoman Lu, Ganqu Cui,
  Taiwen He, Zhiyuan Liu, Tat-Seng Chua, and Maosong Sun. 2024{\natexlab{b}}.
\newblock Rlaif-v: Aligning mllms through open-source ai feedback for super
  gpt-4v trustworthiness.
\newblock \emph{arXiv preprint arXiv:2405.17220}.

\bibitem[{Zhang et~al.(2024)Zhang, Singh, Liu, Liu, Yu, Gao, and
  Zhao}]{zhang2024pasta}
Qingru Zhang, Chandan Singh, Liyuan Liu, Xiaodong Liu, Bin Yu, Jianfeng Gao,
  and Tuo Zhao. 2024.
\newblock Tell your model where to attend: Post-hoc attention steering for
  {LLM}s.
\newblock In \emph{The Twelfth International Conference on Learning
  Representations (ICLR)}.

\bibitem[{Zheng et~al.(2024)Zheng, Wang, Huang, Song, Tang, Xiong, and
  Li}]{zheng2024attentionheadsurvey}
Zifan Zheng, Yezhaohui Wang, Yuxin Huang, Shichao Song, Bo~Tang, Feiyu Xiong,
  and Zhiyu Li. 2024.
\newblock Attention heads of large language models: A survey.
\newblock \emph{arXiv preprint arXiv:2409.03752}.

\bibitem[{Zhou et~al.(2024)Zhou, Cui, Yoon, Zhang, Deng, Finn, Bansal, and
  Yao}]{lure}
Yiyang Zhou, Chenhang Cui, Jaehong Yoon, Linjun Zhang, Zhun Deng, Chelsea Finn,
  Mohit Bansal, and Huaxiu Yao. 2024.
\newblock Analyzing and mitigating object hallucination in large
  vision-language models.
\newblock In \emph{Proceedings of the International Conference on Learning
  Representations (ICLR)}.

\bibitem[{Zhu et~al.(2023)Zhu, Chen, Shen, Li, and Elhoseiny}]{minigpt4}
Deyao Zhu, Jun Chen, Xiaoqian Shen, Xiang Li, and Mohamed Elhoseiny. 2023.
\newblock Minigpt-4: Enhancing vision-language understanding with advanced
  large language models.
\newblock \emph{arXiv preprint arXiv:2304.10592}.

\end{thebibliography}

\clearpage

\appendix
\crefalias{section}{appendix}
\crefalias{subsection}{appendix}
\crefalias{subsubsection}{appendix}

\startcontents[appendix]

{ 
  \printcontents[appendix]{ }{0}{\section*{Appendix}}
} 

\newpage

\begin{figure*}[t!]
    \centering
    \includegraphics[width=0.9\linewidth]{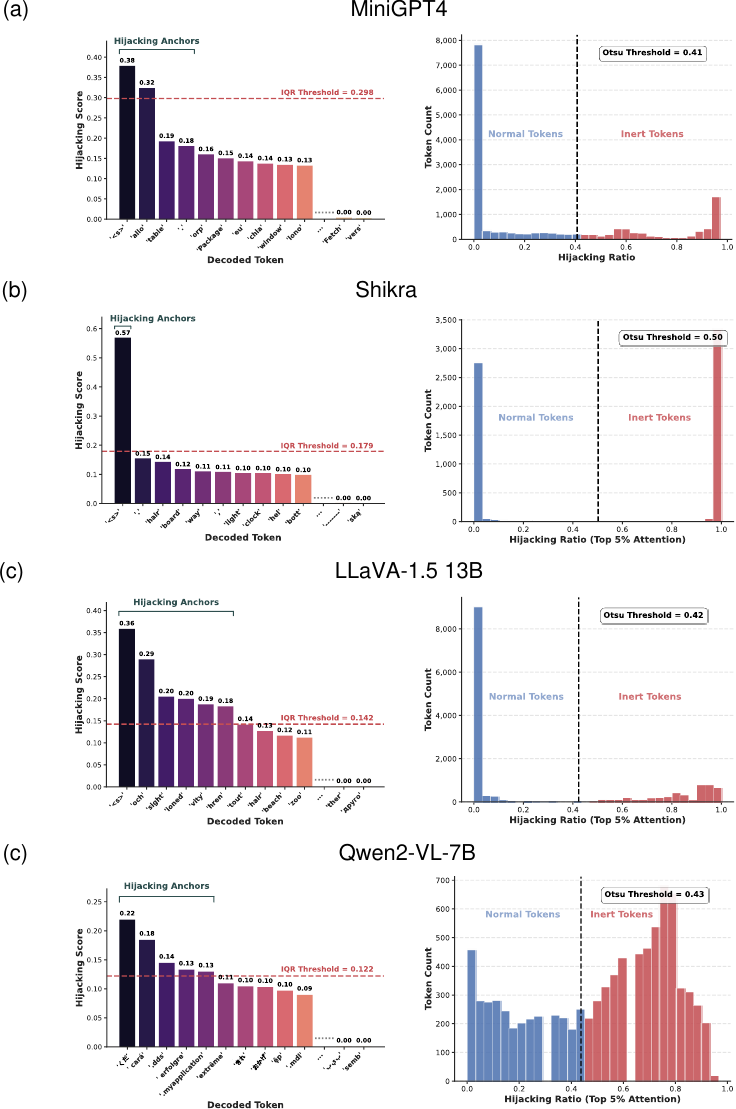}
    \caption{
    Validation of the HABI identification strategy across diverse LVLM architectures. 
    Consistent with the findings on LLaVA-1.5 7B, all tested models exhibit the characteristic long-tailed distribution of Hijacking Scores (enabling \textbf{Hijacking Anchor} identification) and the bimodal distribution of Hijacking Ratios (facilitating \textbf{Inert Token} isolation). 
    The figure displays results for: 
    \textbf{(a)} MiniGPT-4, 
    \textbf{(b)} Shikra, 
    \textbf{(c)} LLaVA-1.5 13B, and 
    \textbf{(d)} Qwen2-VL-7B.
    }
    \label{fig:threshold_app}
\end{figure*}

\section{The Use of Large Language Models}
Throughout the preparation of this manuscript, large language models were employed exclusively for light stylistic refinement and the occasional grammatical adjustment. Every conceptual insight, analytical thread, and interpretive conclusion emerged from the authors themselves; no algorithmic assistance was solicited for the framing, design, or substance of the work, and full scientific responsibility rests with the human contributors alone.

\begin{figure*}[t]
    \centering
    \includegraphics[width=\linewidth]{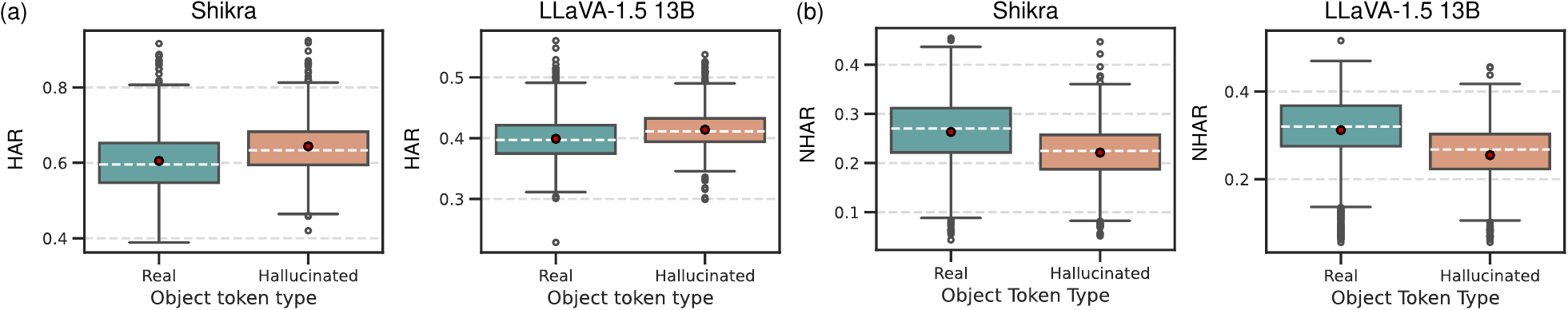}
    \caption{
        Supporting analysis of metric distributions on additional models. 
        \textbf{(a)} HAR distributions for Shikra and LLaVA-1.5 13B. 
        \textbf{(b)} NHAR distributions for the same models.
        }
    \label{fig:HAR_NHAR_app}
\end{figure*}

\begin{figure*}[t]
    \centering
    \includegraphics[width=\linewidth]{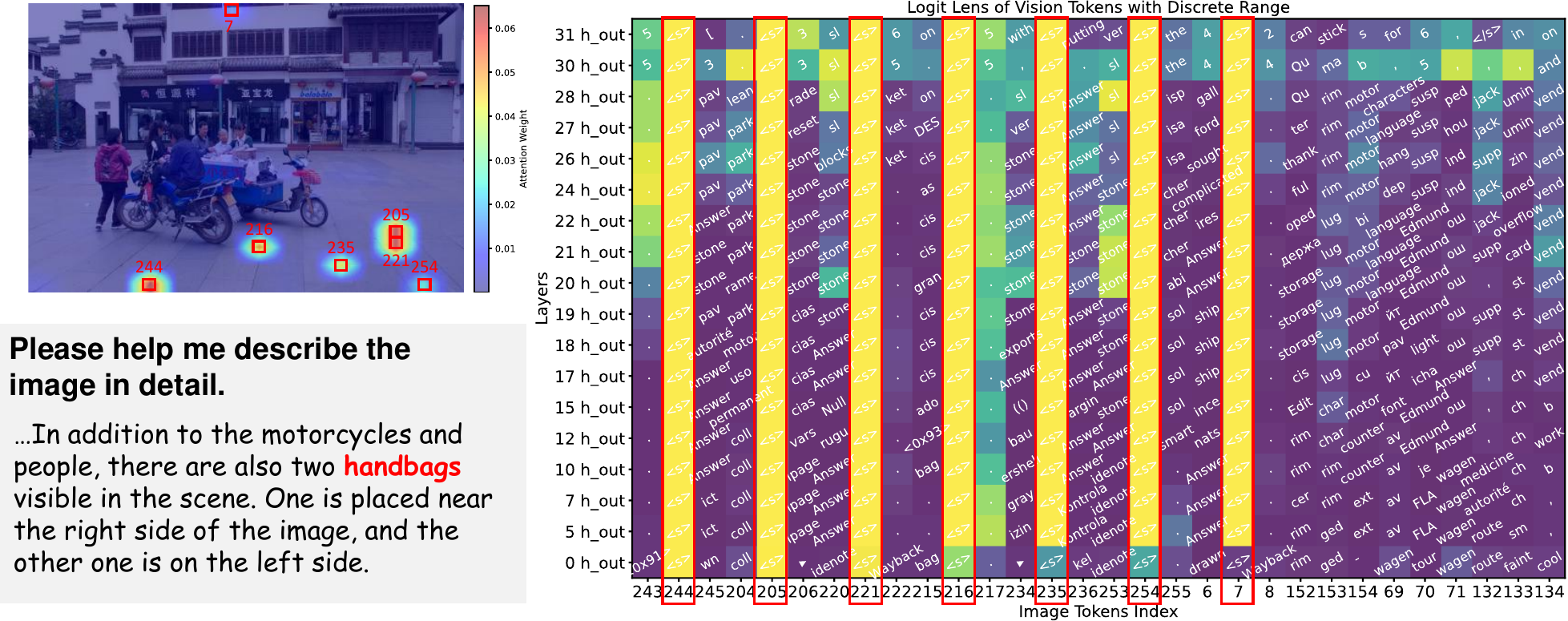}
    \caption{
        Visualization of Vocabulary Hijacking in \textbf{Shikra}. 
        Following the visualization protocol established in \Cref{fig:intro}, this figure demonstrates a significantly more severe hijacking phenomenon: the vast majority of the model's attention is captured by \textbf{Inert Tokens}. 
        This visual evidence aligns with the elevated HAR scores observed in \Cref{app:nvar_vasr}, explaining Shikra's higher susceptibility to hallucinations.
    }
    \label{fig:VH_shikra}
\end{figure*}

\begin{figure*}[t]
    \centering
    \includegraphics[width=\linewidth]{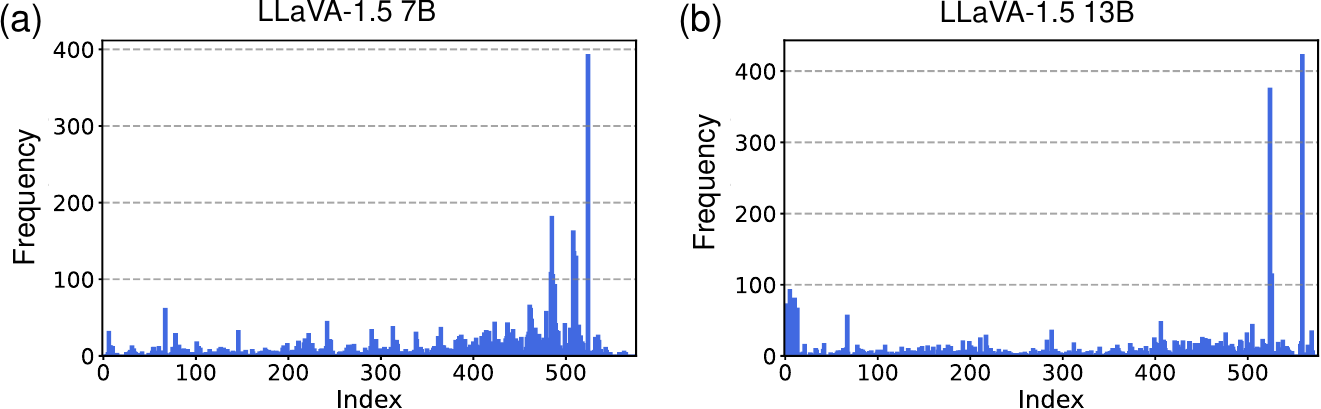}
     \caption{Positional distribution of Inert tokens for LLaVA-1.5 7B and 13B within the 576-token visual sequence. The plots show that Inert tokens are not uniformly distributed, but are instead highly concentrated in specific index ranges. This suggests that the Vocabulary Hijacking phenomenon may be intrinsically linked to the mechanics of the attention mechanism.}
    \label{fig:VAS_index}
\end{figure*}

\begin{figure}[t]
    \centering
    \includegraphics[width=\linewidth]{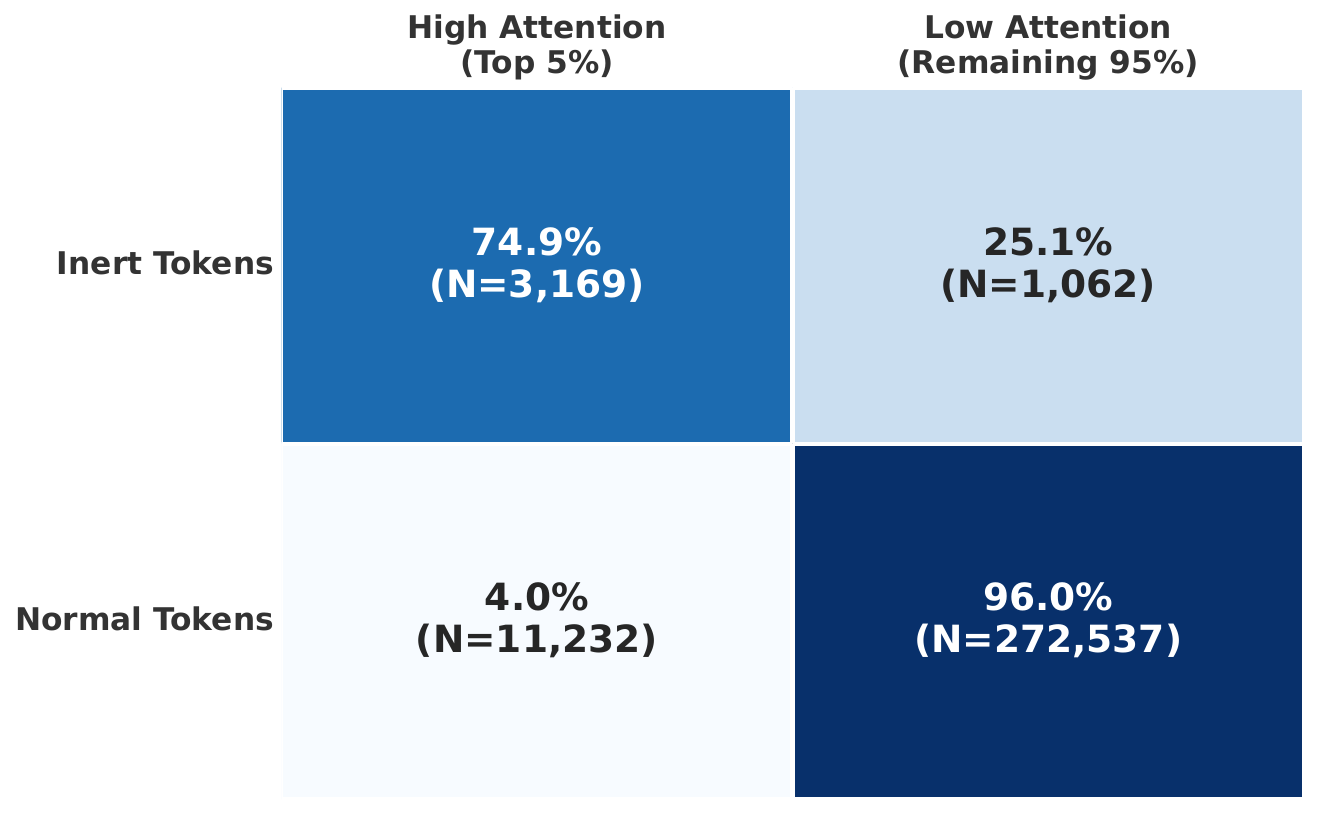}
    \caption{\textbf{Conditional Probability of Attention Distribution.} 
    Although Inert Tokens (top row) constitute only $\sim1\%$ of the total visual tokens, over 75\% of them are concentrated in the High Attention (Top 5\%) region. 
    In contrast, Normal Tokens (bottom row) follow a standard distribution. 
    This confirms the hijacking nature of Inert Tokens, demonstrating their capacity to actively capture the model's focus.}
    \label{fig:inert_attention}
\end{figure}

\section{More Statistical Results}

\subsection{Empirical Results for HABI on Additional Models}
\label{app:HABI}

This section validates the universality of our HABI identification strategy across diverse architectures, including MiniGPT-4, Shikra, LLaVA-1.5 13B, and Qwen2-VL-7B. 
As visualized in \Cref{fig:threshold_app}, all models exhibit statistical behaviors consistent with LLaVA-1.5 7B: a \textbf{long-tailed distribution} of Hijacking Scores and a distinct \textbf{bimodal distribution} of Hijacking Ratios among salient tokens. 
These ubiquitous patterns confirm that the Vocabulary Hijacking phenomenon is intrinsic to LVLMs, thereby justifying the robustness of our statistical thresholds for reliably isolating \textbf{Inert Tokens}. 

We note specific adaptation details for distinct architectures: due to the distribution characteristics of Qwen2-VL-7B, we employed the third quartile ($Q3$) as the threshold for identifying Hijacking Anchors, rather than the standard $Q3 + 1.5 \times \text{IQR}$. 
Additionally, for MiniGPT-4, given its limited visual sequence length (only 32 tokens), we calculated the Hijacking Ratio using the entire set of visual tokens, bypassing the top 5\% attention filtering.

\subsection{NHAR and HAR Distributions on Additional Models}
\label{app:nvar_vasr}

\Cref{fig:HAR_NHAR_app} (a) illustrates the HAR distributions for LLaVA-1.5 7B and 13B, confirming that hallucinated tokens are consistently associated with elevated HAR scores. 
By synthesizing these results with those in the main text (\Cref{fig:HAR_NHAR}), we observe the varying severity of Vocabulary Hijacking across the tested models: Shikra exhibits the most pronounced susceptibility, while LLaVA-1.5 7B shows the mildest effect.

To substantiate this observation, we present a visualization of Vocabulary Hijacking on Shikra in \Cref{fig:VH_shikra}. 
The attention map reveals that a predominant portion of the model's focus is captured by \textbf{Inert Tokens}. 
This visual evidence corroborates the statistical trend of higher HAR scores and suggests a potential mechanism underlying Shikra's inferior performance relative to the LLaVA series.

\Cref{fig:HAR_NHAR_app} (b) presents the NHAR distributions for Shikra and LLaVA-1.5 13B. 
Aligning with our primary findings, real object tokens yield markedly higher NHAR scores compared to hallucinated tokens across these architectures. 
This consistent distinction further validates the robustness and generalizability of NHAR as a reliable criterion for identifying \textbf{hallucination-critical attention heads}.

\subsection{Positional Distribution of Inert Tokens}
\label{app:VAS_idnex}

To further investigate the properties of \textbf{Inert Tokens}, we analyze their positional distribution within the visual sequence. 
As shown in \Cref{fig:VAS_index}, we present the statistical distribution of the relative indices (from 0 to 575) for \textbf{Inert Tokens} in LLaVA-1.5 7B and 13B. 
The results reveal that \textbf{Inert Tokens} tend to be highly concentrated in specific index ranges, rather than being uniformly distributed. 
This positional bias suggests that the emergence of \textbf{Vocabulary Hijacking} may be intrinsically linked to the underlying dynamics of the attention mechanism itself.

\subsection{Quantifying the Attention Attraction of Inert Tokens}
\label{app:inert_attention_analysis}

To empirically validate whether the semantically rigid tokens identified by our method (i.e., \textbf{Inert Tokens}) indeed act as attention sinks, we analyze their attention weight distributions in \Cref{fig:inert_attention}. 
We categorize visual tokens into Inert and Normal groups, further classifying them based on attention magnitude (defining ``High Attention'' as the top 5\% percentile).

A striking disparity is observed: while Inert Tokens constitute a minute fraction of the total sequence ($\approx 1\%$), a staggering 75\% of them fall into the High Attention category. 
In contrast, the vast majority of Normal Tokens exhibit low attention weights. 
This disproportionate concentration confirms our core motivation: Inert Tokens are not merely semantically empty; they are the primary drivers of attention hijacking, actively draining the model's focus despite their scarcity.

\begin{figure*}[t]
    \centering
    \includegraphics[width=\linewidth]{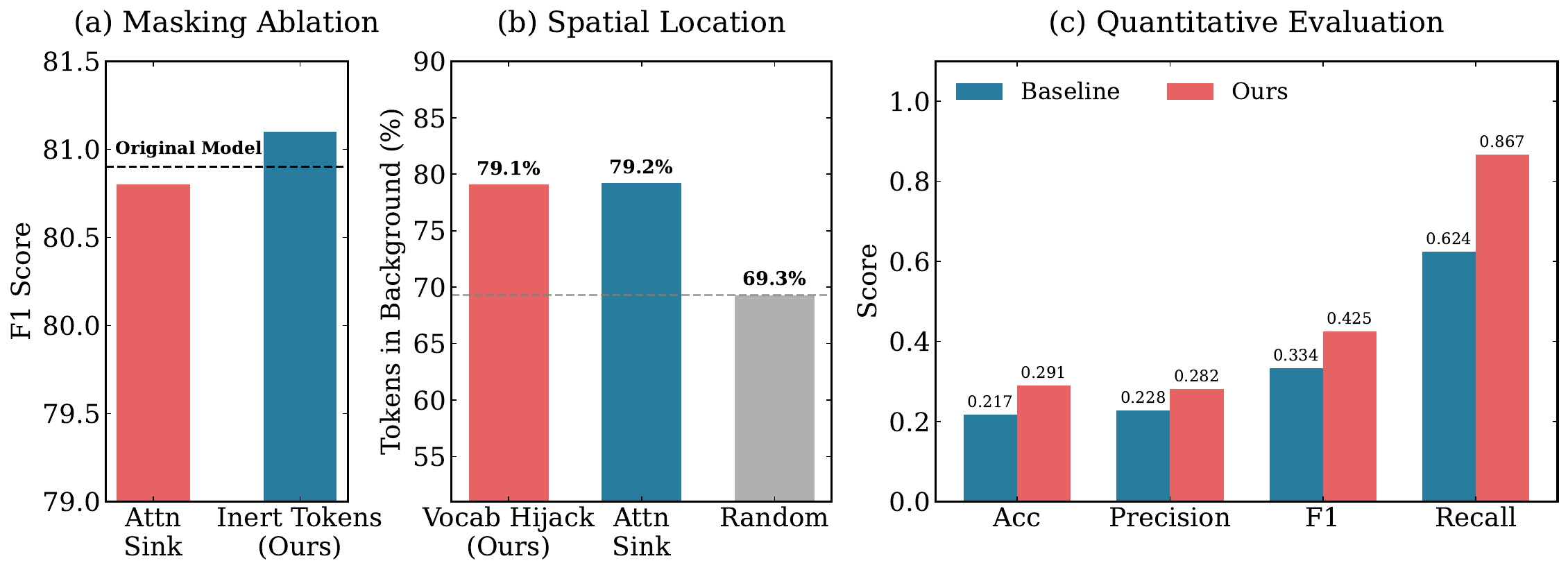}
    \caption{\textbf{Comparative analysis of attention anomaly identification methods.} 
    \textbf{(a) Zero-Ablation Verification:} Changes in POPE F1 scores when masking out specific token sets (Massive Activation vs. HABI), used to quantify the functional redundancy of the identified tokens. 
    \textbf{(b) Spatial Alignment:} The proportion of tokens falling within image background regions (defined by segmentation masks) across different selection methods (Random, Massive Activation, HABI). 
    \textbf{(c) Identification Accuracy:} Evaluation of the alignment with the behavioral ground truth ($\mathcal{G}_{\text{persist}}$). The chart reports Precision, Recall, and F1 scores, measuring the capability of each method to retrieve tokens that exhibit persistent attention patterns.}
    \label{fig:Comparison_VAS}
\end{figure*}

\begin{figure*}[t]
    \centering
    \includegraphics[width=0.9\linewidth]{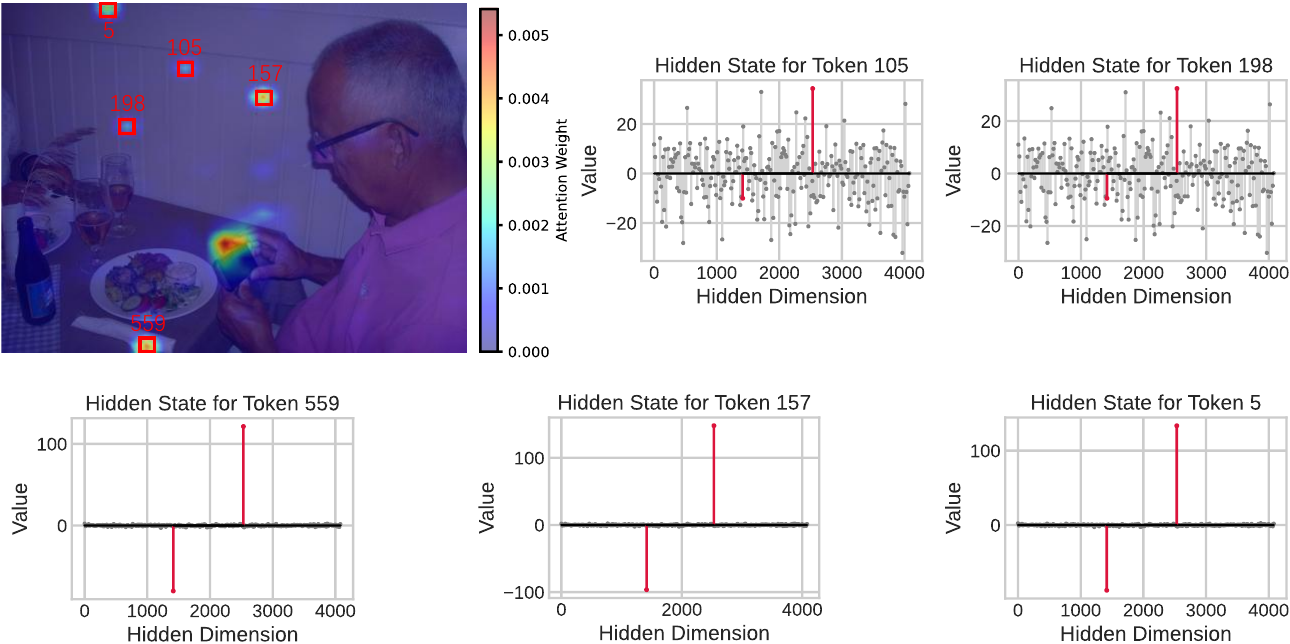}
    \caption{
        \textbf{Visualization of hidden state distributions for five identified Inert Tokens at layer 5.} 
        The red vertical lines indicate the specific dimensions (1415 and 2533) used by the Massive Activation baseline to detect anomalies. 
        Crucially, while some tokens exhibit the expected spikes, \textbf{tokens 105 and 198 do not show massive activation in the designated dimensions}. 
        This absence demonstrates why the activation-based method fails to detect these potent hijackers, whereas our HABI method successfully identifies them.
    }
    \label{fig:vfi_compare}
\end{figure*}

\section{Comparison with Visual Attention Sink}
\label{app:vas_comparison}

In this section, we clarify the relationship between Vocabulary Hijacking and the Visual Attention Sink (VAS) phenomenon. 
While both share certain behavioral characteristics, we demonstrate that our proposed method offers a more precise identification of underlying attention anomalies compared to the activation-based method~\citep{visattnsink}.

\subsection{Verification of General Characteristics}

We first verify that the \textbf{Inert Tokens} identified by HABI share the defining characteristics typically associated with the Visual Attention Sink (VAS) phenomenon: negligible functional contribution and background localization.

\paragraph{Low Functional Contribution (Zero-Ablation Study).} 
To assess the functional importance of these tokens, we conducted a zero-ablation experiment on a subset of the POPE benchmark. 
As shown in~\Cref{fig:Comparison_VAS} (a), masking out either the \textbf{Inert Tokens} (identified by HABI) or the VAS tokens (identified by Massive Activation) has a negligible impact on the F1 score.
This outcome confirms that both token categories correspond to informationally sparse regions that are functionally redundant for the generation task.

\paragraph{Background Localization.} 
We quantitatively analyze the spatial distribution of these tokens relative to semantic regions. 
By utilizing ground-truth segmentation masks, we define the image background as the complementary area to all annotated objects. 
As visualized in~\Cref{fig:Comparison_VAS} (b), both \textbf{Inert Tokens} and VAS tokens exhibit a significantly higher probability of residing in background regions compared to a random baseline. 
This empirical evidence confirms that both token types are predominantly localized within the background.

\subsection{Analysis of Critical Properties}
\label{sec:eval_persistence}

While background localization and functional redundancy are observable symptoms, we argue that they are merely \textit{secondary} characteristics. 
The \textbf{definitive feature} of this phenomenon is \textbf{Persistent Attention Attraction}—the capacity of specific tokens to consistently monopolize attention across multiple generation steps, disregarding the evolving context.

To quantitatively evaluate which identification paradigm better captures this core anomaly, we constructed a behavior-based ground truth and performed a comparative analysis between our \textbf{Hijacking Anchor-Based Identification (HABI)} and the \textbf{Massive Activation method} used in prior work~\citep{visattnsink}.

\paragraph{Constructing the Behavioral Ground Truth.}
Given the absence of explicit annotations for attention anomalies, we establish a behavioral approximation of the ground truth. 
\textbf{Strictly speaking, this constructed set represents a subset of the true ground truth.}
We hypothesize that a true hijacking token must consistently attract high attention during the initial generation phase. 
For each image, we define the \textbf{Persistent Attention Set}, denoted as $\mathcal{G}_{\text{persist}}$, by computing the intersection of the top-$K$ attended vision tokens across the first $T$ decoding steps. 
Formally, let $\mathcal{S}_t$ be the set of indices for vision tokens with the top-$K$ attention weights at step $t$. The ground truth is defined as:
\begin{equation}
    \mathcal{G}_{\text{persist}} = \bigcap_{t=1}^{T} \mathcal{S}_t.
\end{equation}
In our experiments, we set $T=5$ and $K=10$, aiming to capture tokens that act as stable attention sinks during the critical initial captioning phase.

\paragraph{Comparative Results.}
We evaluate two identification methods against $\mathcal{G}_{\text{persist}}$:
\begin{enumerate}
    \item \textbf{Massive Activation (Baseline):} Identifies tokens where specific dimensions of the hidden states exhibit anomalously high activation values~\citep{visattnsink}.
    \item \textbf{HABI (Ours):} Identifies \textbf{Inert Tokens} using the proposed Hijacking Ratio thresholding ($r_{\text{hijack}} > \tau_r$).
\end{enumerate}

We calculate the Accuracy, Precision, Recall, and F1-score of both methods in retrieving the tokens in $\mathcal{G}_{\text{persist}}$. 
As shown in~\Cref{fig:Comparison_VAS} (c), our HABI method significantly outperforms the Massive Activation baseline across all evaluated metrics.

Crucially, to quantify the distinctness of the identified anomalies, we highlight that HABI achieves a Unique Ratio of 42.93\%, meaning that nearly half of the Inert Tokens we identify are completely overlooked by the activation-based baseline.

This substantial divergence suggests that \textbf{activation magnitude is an insufficient proxy for attention anomalies}. 
While many sink tokens do possess large activations, a substantial portion of ``hijackers'' (which persistently attract attention) do not exhibit massive norms but are accurately detected by our HABI. 
Thus, HABI provides a more robust and mechanically grounded identification of the attention hijacking phenomenon.

\paragraph{Case Study: Failure of the Activation-based Method.}
To illustrate the limitation of the baseline, we revisit the example from \Cref{fig:intro}, where the model generated the token ``phone''. 
Among the top-10 most attended visual tokens, five (indices 5, 105, 157, 198, and 559) were identified as clear persistent sinks. 
We visualize the layer-5 hidden state distributions for these five tokens in \Cref{fig:vfi_compare}, specifically highlighting the massive activation dimensions $\mathcal{D}_{\text{sink}}=\{1415, 2533\}$ identified by \citep{visattnsink}.
The visualization reveals a critical discrepancy: while several tokens exhibit the massive activation pattern, \textbf{tokens 105 and 198 do not show massive activation in the designated dimensions}. 

Consequently, the activation-based method fails to identify these two potent attention hijackers. 
In contrast, HABI successfully flags them due to their rigid Traces, demonstrating superior sensitivity to the phenomenon's true nature.

\section{Comparison with Concurrent Work: AVISC}
\label{app:avisc_comparison}

Following our analysis distinguishing Vocabulary Hijacking from the Visual Attention Sink (VAS) phenomenon in \Cref{app:vas_comparison}, we now clarify the relationship between our method and a concurrent work, AVISC \citep{AVISC}. While both works aim to mitigate LVLM hallucinations, our concept of \textit{Vocabulary Hijacking} (driven by Inert Tokens) and AVISC's \textit{Blind Tokens} represent fundamentally distinct phenomena. We elaborate on these differences from two primary perspectives: mechanistic behavior and empirical performance.

\subsection{Fundamental Mechanistic Differences}
Beyond the divergence from VAS discussed previously, quantitative set analysis demonstrates that AVISC's Blind Tokens and our Inert Tokens capture entirely disjoint subsets of visual features.

\paragraph{Set Divergence.} 
We expanded our set divergence analysis to include Blind Tokens, measuring the overlap between different token sets via Jaccard Distance and Unique Ratio. As reported in \Cref{tab:set_divergence}, the remarkably high Jaccard Distance ($0.8302$) between Inert Tokens and Blind Tokens confirms that they represent highly distinct subsets. Furthermore, the distance between VAS (Attn Sink) and Blind Tokens is even higher ($0.8852$), indicating that Blind Tokens are distinct from both Vocabulary Hijacking and the attention sink phenomenon.

\begin{table}[t]
\centering
\resizebox{\linewidth}{!}{%
\begin{tabular}{l|cc}
\toprule
\textbf{Comparison Pair} & \textbf{Jaccard Dist} & \textbf{Unique Ratio} \\
\midrule
Inert Tokens vs. Attn Sink    & 0.5596 & 0.4291 \\
Inert Tokens vs. Blind Tokens & 0.8302 & 0.0926 \\
Attn Sink vs. Blind Tokens    & 0.8852 & 0.2942 \\
\bottomrule
\end{tabular}
}
\caption{Three-Way Set Divergence Analysis. A higher Jaccard Distance indicates less overlap between the identified token sets.}
\label{tab:set_divergence}
\end{table}

\paragraph{Background Distribution Discrepancy.} 
In \Cref{app:vas_comparison}, we empirically established that both Inert Tokens and VAS tokens are predominantly localized within the image background (approximately $79\%$ for both). In stark contrast, as detailed in \Cref{tab:bg_distribution}, the background ratio of Blind Tokens ($69.75\%$) is merely on par with that of randomly sampled tokens ($68.99\%$). Because Blind Tokens are heuristically defined by simple attention-score thresholding, they are prone to mistakenly selecting foreground semantic targets. Conversely, Inert Tokens represent a highly specific, mechanistically driven semantic collapse firmly rooted in non-semantic background regions.

\begin{table}[t]
\centering
\resizebox{\linewidth}{!}{%
\begin{tabular}{l|ccc}
\toprule
\textbf{Token Type} & \textbf{Total Count} & \textbf{In Background} & \textbf{BG Ratio (\%)} \\
\midrule
Inert Tokens & 16,438 & 12,972 & 78.91 \\
Attn Sink    & 15,030 & 11,903 & 79.19 \\
Blind Tokens & 8,785  & 6,127  & 69.75 \\
Random       & 20,000 & 13,798 & 68.99 \\
\bottomrule
\end{tabular}
}
\caption{Background Distribution Analysis. Unlike Inert Tokens and Attention Sinks, which show a strong tendency to concentrate in the visual background, Blind Tokens align closely with a random spatial distribution.}
\label{tab:bg_distribution}
\end{table}

\subsection{Empirical Superiority and Computational Efficiency}
The precision of our mechanistic identification directly translates into performance and efficiency gains. As shown in \Cref{tab:comprehensive_benchmark}, our HAVAE method consistently outperforms AVISC across standard hallucination benchmarks, including POPE, MME, and AMBER. 

To ensure a rigorous and fair setup during our evaluation, we set the maximum generation length to 512 tokens for AVISC, which accounts for the numerical differences compared to their original reported results. Despite this standardized setting, HAVAE consistently achieves higher Accuracy and F1 scores. 

Crucially, HAVAE maintains a significantly lower computational footprint. Unlike AVISC, which relies on contrastive decoding and thus requires an additional forward pass, HAVAE is a single-pass intervention that selectively reinforces attention heads. Consequently, our method achieves superior hallucination mitigation while consuming approximately half the inference computing resources required by AVISC.

\begin{table*}[t]
    \centering
    \resizebox{0.9\textwidth}{!}{%
    \begin{tabular}{l|cc|cc|ccc}
        \toprule
        \multicolumn{1}{l}{\multirow{2}{*}{\textbf{Method}}}
        & \multicolumn{2}{c}{\textbf{POPE}} 
        & \multicolumn{2}{c}{\textbf{MME}} 
        & \multicolumn{3}{c}{\textbf{AMBER}} \\
        \cmidrule(lr){2-3} \cmidrule(lr){4-5} \cmidrule(lr){6-8}
        & \textbf{Acc. $\uparrow$} & \textbf{F1 $\uparrow$} 
        & \textbf{Per. $\uparrow$} & \textbf{Cog. $\uparrow$} 
        & \textbf{CHAIR $\downarrow$} & \textbf{Acc. $\uparrow$} & \textbf{F1 $\uparrow$} \\
        \midrule
        AVISC & 83.2 & 84.1 & 1379.3 & 321.4 & 12.6 & 70.7 & 75.2 \\
        \textbf{HAVAE(Ours)} & \textbf{86.2} & \textbf{86.3} & \textbf{1483.8} & \textbf{327.9} & \textbf{3.6} & \textbf{78.6} & \textbf{82.7} \\
        \bottomrule
    \end{tabular}
    }
    \caption{Comprehensive performance comparison between AVISC and our proposed HAVAE across standard hallucination benchmarks.}
    \label{tab:comprehensive_benchmark}
\end{table*}

\begin{table*}[t]
\centering

\resizebox{0.95\textwidth}{!}{%
\begin{tabular}{ll|ccc|cc|cc}
\toprule
\multicolumn{1}{l}{\multirow{2}{*}{\textbf{Model}}}
&\multicolumn{1}{l}{\multirow{2}{*}{\textbf{Method}}}
& \multicolumn{3}{c}{\textbf{CHAIR}} 
& \multicolumn{2}{c}{\textbf{POPE}} 
& \multicolumn{2}{c}{\textbf{POPE Chat}} \\ 
\cmidrule(lr){3-5} \cmidrule(lr){6-7} \cmidrule(lr){8-9}
& & \textbf{CHAIR$_s$ $\downarrow$} & \textbf{CHAIR$_i$ $\downarrow$}& \textbf{F1 $\uparrow$} & \textbf{Acc. $\uparrow$} & \textbf{F1 $\uparrow$} & \textbf{Acc. $\uparrow$} & \textbf{F1 $\uparrow$} \\
\midrule
\multicolumn{1}{l}{\multirow{5}{*}{LLaVA-1.5-7B}} & Beam Search        & 47.6 & 13.0 & \cellcolor{gray!15} 79.0 & 84.7 & 85.4 & 85.3 & 83.2 \\
& PAI               & 21.6 & 6.2  & \cellcolor{gray!15} 75.8 & 85.1 & 85.7 & 88.1 & 87.0 \\
& Devils            & 29.0 & 6.8  & \cellcolor{gray!15} 80.1 & 84.9 & 85.6 & \textbf{88.4} & \textbf{87.6} \\
& VISTA    & 9.8  & 5.7  & \cellcolor{mypink}54.3 & 83.1 & 84.6 & --- & --- \\
& \textbf{HAVAE(Ours)} & \textbf{20.0}$^{\textcolor{mygreen}{\scriptsize -7.4\%}}$ & \textbf{5.8}$^{\textcolor{mygreen}{\scriptsize -6.5\%}}$ & \cellcolor{gray!15} 78.7 & \textbf{86.0}$^{\textcolor{mygreen}{\scriptsize +1.1\%}}$ & \textbf{86.2}$^{\textcolor{mygreen}{\scriptsize +0.6\%}}$ & 88.1$^{\textcolor{myred}{\scriptsize -0.3\%}}$ & 87.1$^{\textcolor{myred}{\scriptsize -0.6\%}}$ \\
\midrule
\multicolumn{1}{l}{\multirow{5}{*}{MiniGPT-4-7B}} & Beam Search    & 29.0 & 8.9 & \cellcolor{gray!15} 72.9 & 76.9 & 76.9 & 77.9 & 78.0 \\
& PAI               & 23.0 & 7.5 & \cellcolor{gray!15} 72.8 & 75.5 & 76.9 & 78.5 & 78.6 \\
& Devils            & 20.2 & 6.9 & \cellcolor{gray!15} 71.9 & 67.2 & 74.3 & 79.3 & 79.5 \\
& VISTA    & 15.4 & 4.6 & \cellcolor{mypink}67.4 & 76.3 & 77.2 & ---  & ---  \\
& \textbf{HAVAE(Ours)} & \textbf{20.0}$^{\textcolor{mygreen}{\scriptsize -1.0\%}}$ & \textbf{6.6}$^{\textcolor{mygreen}{\scriptsize -4.3\%}}$ & \cellcolor{gray!15} 74.3 & \textbf{77.1}$^{\textcolor{mygreen}{\scriptsize +0.3\%}}$ & \textbf{78.1}$^{\textcolor{mygreen}{\scriptsize +1.2\%}}$ & \textbf{79.7}$^{\textcolor{mygreen}{\scriptsize +0.5\%}}$ & \textbf{80.4}$^{\textcolor{mygreen}{\scriptsize +1.1\%}}$ \\
\midrule
\multicolumn{1}{l}{\multirow{5}{*}{Shikra-7B}} & Beam Search        & 56.6 & 14.1 & \cellcolor{gray!15} 77.0 & 81.1 & 81.6 & 77.3 & 78.5 \\
& PAI               & 35.4 & 9.2  & \cellcolor{gray!15} 77.1 & \textbf{82.0} & 81.3 & 77.4 & 77.3 \\
& Devils            & 21.2 & 8.1  & \cellcolor{gray!15} 73.6 & 80.3 & 80.5 & 76.9 & 78.0 \\
& VISTA    & 31.4 & 10.7 & \cellcolor{gray!15} 74.4 & 81.3 & 82.1 & ---  & ---  \\
& \textbf{HAVAE(Ours)} & \textbf{15.0}$^{\textcolor{mygreen}{\scriptsize -29.2\%}}$ & \textbf{3.4}$^{\textcolor{mygreen}{\scriptsize -58.0\%}}$ & \cellcolor{gray!15} 72.2 & 81.7$^{\textcolor{myred}{\scriptsize -0.4\%}}$ & \textbf{82.2}$^{\textcolor{mygreen}{\scriptsize +0.1\%}}$ & \textbf{77.5}$^{\textcolor{mygreen}{\scriptsize +0.1\%}}$ & \textbf{78.9}$^{\textcolor{mygreen}{\scriptsize +0.5\%}}$ \\
\bottomrule
\end{tabular}
}
\caption{
Performance of \textbf{HAVAE} against baselines using beam search decoding. Best results are in \textbf{bold}. Pink cells mark potentially unreliable CHAIR scores. Superscripts show the \% change vs. the best baseline. 
}
\label{tab:main_results_beam}
\end{table*}
\begin{table*}[t]
\centering

\resizebox{0.95\textwidth}{!}{%
\begin{tabular}{ll|ccc|cc|cc}
\toprule
\multicolumn{1}{l}{\multirow{2}{*}{\textbf{Model}}}
&\multicolumn{1}{l}{\multirow{2}{*}{\textbf{Method}}}
& \multicolumn{3}{c}{\textbf{CHAIR}} 
& \multicolumn{2}{c}{\textbf{POPE}} 
& \multicolumn{2}{c}{\textbf{POPE Chat}} \\ 
\cmidrule(lr){3-5} \cmidrule(lr){6-7} \cmidrule(lr){8-9}
& & \textbf{CHAIR$_s$ $\downarrow$} & \textbf{CHAIR$_i$ $\downarrow$}& \textbf{F1 $\uparrow$} & \textbf{Acc. $\uparrow$} & \textbf{F1 $\uparrow$} & \textbf{Acc. $\uparrow$} & \textbf{F1 $\uparrow$} \\
\midrule
\multicolumn{1}{l}{\multirow{5}{*}{LLaVA-1.5-7B}} & Sample        & 48.2 & 15.2 & \cellcolor{gray!15} 73.8 & 83.2 & 84.0 & 85.1 & 83.1 \\
& PAI               & 41.6 & 11.3 & \cellcolor{gray!15} 71.7 & 83.5 & 84.2 & 87.0 & 85.9 \\
& Devils            & 31.8 & 7.1 & \cellcolor{gray!15} 79.9 & 83.7 & \textbf{84.3} & \textbf{87.3} & \textbf{86.1} \\
& VISTA    & 16.0 & 7.9  & \cellcolor{mypink}65.7 & 82.6 & 84.0 & --- & --- \\
& \textbf{HAVAE(Ours)} & \textbf{24.8}$^{\textcolor{mygreen}{\scriptsize -22.0\%}}$ & \textbf{5.5}$^{\textcolor{mygreen}{\scriptsize -22.5\%}}$ & \cellcolor{gray!15} 77.2 & \textbf{84.0}$^{\textcolor{mygreen}{\scriptsize +0.4\%}}$ & \textbf{84.3}$^{\textcolor{mygreen}{\scriptsize \pm0.0\%}}$ & 87.0$^{\textcolor{myred}{\scriptsize -0.3\%}}$ & \textbf{86.1}$^{\textcolor{mygreen}{\scriptsize \pm0.0\%}}$ \\
\midrule
\multicolumn{1}{l}{\multirow{5}{*}{MiniGPT-4-7B}} & Sample    & 33.8 & 10.4 & \cellcolor{gray!15} 71.4 & 67.2 & 68.2 & 74.2 & 74.2 \\
& PAI               & 28.4 & 12.2 & \cellcolor{gray!15} 69.1 & 65.9 & 68.9 & 75.4 & 75.4 \\
& Devils            & 22.2 & \textbf{7.7} & \cellcolor{gray!15} 71.9 & 63.0 & 68.0 & 75.0 & 74.5 \\
& VISTA    & 17.4 & 4.8  & \cellcolor{mypink}67.7 & 66.9 & 68.1 & ---  & ---  \\
& \textbf{HAVAE(Ours)} & \textbf{22.0}$^{\textcolor{mygreen}{\scriptsize -0.9\%}}$ & 7.8$^{\textcolor{myred}{\scriptsize +1.3\%}}$ & \cellcolor{gray!15} 72.9 & \textbf{67.4}$^{\textcolor{mygreen}{\scriptsize +0.3\%}}$ & \textbf{69.1}$^{\textcolor{mygreen}{\scriptsize +0.3\%}}$ & \textbf{75.7}$^{\textcolor{mygreen}{\scriptsize +0.4\%}}$ & \textbf{76.7}$^{\textcolor{mygreen}{\scriptsize +1.7\%}}$ \\
\midrule
\multicolumn{1}{l}{\multirow{5}{*}{Shikra-7B}} & Sample        & 57.4 & 16.1 & \cellcolor{gray!15} 73.7 & 79.7 & 80.7 & \textbf{75.7} & 77.7 \\
& PAI               & 41.6 & 11.4 & \cellcolor{gray!15} 72.4 & 80.1 & 80.2 & 75.6 & 76.7 \\
& Devils            & 25.0 & 8.8  & \cellcolor{gray!15} 73.3 & 78.6 & 79.4 & 75.2 & 77.3 \\
& VISTA    & 32.6 & 11.0 & \cellcolor{gray!15} 72.5 & \textbf{80.2} & 81.0 & ---  & ---  \\
& \textbf{HAVAE(Ours)} & \textbf{18.2}$^{\textcolor{mygreen}{\scriptsize -27.2\%}}$ & \textbf{4.6}$^{\textcolor{mygreen}{\scriptsize -47.7\%}}$ & \cellcolor{gray!15} 71.3 & 80.1$^{\textcolor{myred}{\scriptsize -0.1\%}}$ & \textbf{81.1}$^{\textcolor{mygreen}{\scriptsize +0.1\%}}$ & 75.5$^{\textcolor{myred}{\scriptsize -0.3\%}}$ & \textbf{78.0}$^{\textcolor{mygreen}{\scriptsize +0.4\%}}$ \\
\bottomrule
\end{tabular}
}

\caption{
Performance of \textbf{HAVAE} against baselines using sample as decoding strategy. Best results are in \textbf{bold}. Pink cells mark potentially unreliable CHAIR scores. Superscripts show the \% change vs. the best baseline. 
}
\label{tab:main_results_sample}
\end{table*}
\begin{table*}[t]
\centering

\resizebox{0.95\textwidth}{!}{%
\begin{tabular}{ll|cccc|cc|c}
\toprule
\multicolumn{1}{l}{\multirow{2}{*}{\textbf{Model}}}
&\multicolumn{1}{l}{\multirow{2}{*}{\textbf{Method}}}
& \multicolumn{4}{c}{\textbf{Generative}} 
& \multicolumn{2}{c}{\textbf{Discriminative}} 
& \multicolumn{1}{c}{\textbf{AMBER}} \\ 
\cmidrule(lr){3-6} \cmidrule(lr){7-8} 
& & \textbf{CHAIR$_i$ $\downarrow$} & \textbf{Cover $\uparrow$}& \textbf{Hal $\downarrow$} & \textbf{Cog$^*$ $\downarrow$} &  \textbf{Acc. $\uparrow$}  & \textbf{F1 $\uparrow$} & \textbf{Score $\uparrow$} \\
\midrule
\multicolumn{1}{l}{\multirow{4}{*}{LLaVA-1.5-7B}} & Greedy        & 6.0 & 50.6 & 27.4 & 2.8 & 74.8 & 77.6 & 85.8 \\
& PAI               & 5.0 & 46.2 & 20.5 & 1.7 & 78.0 & 81.2 & 88.1 \\
& Devils            & 3.8 & 46.0 & 20.7 & \textbf{1.2} & 77.8 & 81.3 & 88.8 \\
& \textbf{HAVAE(Ours)} & \textbf{3.6}$^{\textcolor{mygreen}{\scriptsize -5.3\%}}$ & \textbf{51.7}$^{\textcolor{mygreen}{\scriptsize +2.2\%}}$ & \textbf{20.2}$^{\textcolor{mygreen}{\scriptsize -1.5\%}}$ & 1.3$^{\textcolor{myred}{\scriptsize +8.3\%}}$ & \textbf{78.6}$^{\textcolor{mygreen}{\scriptsize +0.8\%}}$ & \textbf{82.7}$^{\textcolor{mygreen}{\scriptsize +1.7\%}}$ & \textbf{89.6}$^{\textcolor{mygreen}{\scriptsize +0.9\%}}$ \\
\midrule
\multicolumn{1}{l}{\multirow{4}{*}{MiniGPT-4-7B}} & Greedy        & 15.3 & \textbf{63.3} & 65.2 & 11.0 & 64.9 & 65.1 & 74.9 \\
& PAI               & 12.3 & 60.8 & 51.3 & 7.2  & 61.4 & 61.3 & 74.5 \\
& Devils            & 11.5 & 58.8 & 49.2 & 6.4  & 58.0 & 56.4 & 72.5 \\
& \textbf{HAVAE(Ours)} & \textbf{11.2}$^{\textcolor{mygreen}{\scriptsize -2.6\%}}$ & 61.1$^{\textcolor{myred}{\scriptsize -3.5\%}}$ & \textbf{48.9}$^{\textcolor{mygreen}{\scriptsize -0.6\%}}$ & \textbf{6.1}$^{\textcolor{mygreen}{\scriptsize -4.7\%}}$ & \textbf{65.0}$^{\textcolor{mygreen}{\scriptsize +0.2\%}}$ & \textbf{65.2}$^{\textcolor{mygreen}{\scriptsize +0.2\%}}$ & \textbf{77.0}$^{\textcolor{mygreen}{\scriptsize +2.8\%}}$ \\
\midrule
\multicolumn{1}{l}{\multirow{4}{*}{Shikra-7B}} & Greedy        & 11.2 & \textbf{50.9} & 49.7 & 5.6 & 78.5 & 82.1 & 85.5 \\
& PAI               & 7.2  & 49.3 & 34.3 & 3.0 & 78.0 & 82.0 & 87.4 \\
& Devils            & 6.7  & 45.3 & 22.5 & 1.6 & 71.1 & 74.1 & 83.7 \\
& \textbf{HAVAE(Ours)} & \textbf{3.6}$^{\textcolor{mygreen}{\scriptsize -46.3\%}}$ & 48.9$^{\textcolor{myred}{\scriptsize -3.9\%}}$ & \textbf{17.8}$^{\textcolor{mygreen}{\scriptsize -20.9\%}}$ & \textbf{1.0}$^{\textcolor{mygreen}{\scriptsize -37.5\%}}$ & \textbf{78.6}$^{\textcolor{mygreen}{\scriptsize +0.1\%}}$ & \textbf{82.3}$^{\textcolor{mygreen}{\scriptsize +0.2\%}}$ & \textbf{89.4}$^{\textcolor{mygreen}{\scriptsize +2.3\%}}$ \\
\bottomrule
\end{tabular}
}
\caption{
  AMBER benchmark results on LLaVA-1.5-7B, MiniGPT-4-7B, and Shikra-7B. Best results are in \textbf{bold}. Superscripts show the \% change vs. the best baseline.
}
\label{tab:amber_combined}
\end{table*}

\section{Additional Experimental Setups}
\label{app:exp}

\subsection{Detailed Benchmark and Evaluation Metrics}
\label{app:benchmark}
\paragraph{CHAIR~\citep{rohrbach2018object}.}
The Caption Hallucination Assessment with Image Relevance (CHAIR) metric quantifies hallucination in image captions by comparing generated object mentions against a pre-compiled set of ground-truth objects for each image. An object is considered a hallucination if it is mentioned in the caption but is absent from this ground-truth set. It comprises two scores: instance-level ($\mathrm{CHAIR_I}$) and sentence-level ($\mathrm{CHAIR_S}$), calculated as follows:
\begin{gather}
    \mathrm{CHAIR_I} = \frac{|\{\text{hallucinated objects}\}|}{|\{\text{all mentioned objects}\}|}, \\
    \mathrm{CHAIR_S} = \frac{|\{\text{captions with hallucinated objects}\}|}{|\{\text{all captions}\}|}.
\end{gather}

Our evaluation is conducted on 500 randomly sampled instances from the MSCOCO 2014 validation set. To specifically assess long-form generation, we adopt the setup from PAI\citep{liu2024paying} and Devils\citep{jiang2025devils}, generating descriptions with a `max\_new\_tokens' of 512 using the prompt: ``\texttt{Please help me describe the image in detail.}''.

\paragraph{POPE~\citep{pope}.}
The Polling-based Object Probing Evaluation (POPE) is a benchmark designed within the VQA paradigm to assess object hallucination. It evaluates LVLMs by posing binary questions about object presence, such as ``\texttt{Is there a <object> in the image?}''. The questions are constructed using objects from three distinct sampling strategies to test different aspects of model knowledge: \textit{random} (objects chosen randomly from the dataset), \textit{popular} (frequently occurring objects), and \textit{adversarial} (objects semantically related to those present in the image).
\noindent\textit{Experimental Setup.}
We evaluate on 500 images from the COCO test set, with 6 questions per split for each image, reporting both Accuracy and F1 scores. Furthermore, following~\citep{liu2024paying}, to comprehensively examine performance in conversational contexts, we extend the evaluation to include both single-turn and multi-turn dialogues, a setup we term \textbf{POPE-Chat}.

\paragraph{MME~\citep{bench:mme}.}
The MME benchmark is a comprehensive evaluation suite designed to assess the diverse capabilities of LVLMs. It comprises 14 subtasks organized into two primary capability types: \textit{Perception} (e.g., existence, counting, position, color, OCR) and \textit{Cognition} (e.g., commonsense reasoning, numerical calculation). The benchmark employs a concise Yes/No question format to minimize instruction-following variance and facilitate quantitative analysis.
We conduct evaluations on the full MME dataset. To provide a holistic view of model performance, we report the aggregate scores for the two overarching categories: \textbf{Perception} and \textbf{Cognition}. This split evaluation allows us to verify that our intervention maintains robust performance across both fundamental visual recognition and high-level reasoning tasks.

\paragraph{AMBER~\citep{wang2023llm}.} AMBER is a comprehensive benchmark designed to evaluate multiple facets of hallucination, including object, attribute, and relation errors, across both discriminative and generative tasks. While its discriminative tasks are evaluated using standard metrics (e.g., Accuracy, F1 Score), its generative tasks employ a suite of four specific metrics to assess the quality and faithfulness of model responses. Let $R_{\text{obj}}$ be the set of objects mentioned in the model's response, $G_{\text{obj}}$ be the set of ground-truth objects, and $H_{\text{obj}}$ be a pre-annotated set of common human hallucinations. The generative metrics are defined as follows:
\begin{itemize}
    \item \textbf{CHAIR}: Evaluates the proportion of hallucinated objects among all objects mentioned by the model. Note: This is equivalent to the instance-level CHAIR, and is referred to as $\text{CHAIR}_i$ in our main text.
    \begin{equation}
        \text{CHAIR} = 1 - \frac{|R_{\text{obj}} \cap G_{\text{obj}}|}{|R_{\text{obj}}|}.
    \end{equation}

    \item \textbf{Cover}: Measures the proportion of ground-truth objects that are correctly mentioned in the model's response (i.e., object recall).
    \begin{equation}
        \text{Cover} = \frac{|R_{\text{obj}} \cap G_{\text{obj}}|}{|G_{\text{obj}}|}.
    \end{equation}

    \item \textbf{Hal}: A binary metric that indicates whether any hallucination occurred in the response.
    \begin{equation}
        \text{Hal} = \begin{cases} 1, & \text{if } \text{CHAIR} > 0 \\ 0, & \text{otherwise} \end{cases}.
    \end{equation}

    \item \textbf{Cog}: Assesses the similarity between the model's hallucinations and those common to humans.
    \begin{equation}
        \text{Cog} = \frac{|R_{\text{obj}} \cap H_{\text{obj}}|}{|R_{\text{obj}}|}.
    \end{equation}
\end{itemize}
Finally, to provide a single, unified measure of performance, AMBER also proposes the \textbf{AMBER Score}, which combines the F1 score from discriminative tasks and the CHAIR score from generative tasks:
\begin{equation}
\text{AMBER Score} = \frac{1}{2} \times \left(1 - \text{CHAIR}_i + \text{F1}\right).
\end{equation}
The final reported scores are the average values of these metrics across all queries in the benchmark~\citep{du2025nps,du2025graftllm}.

\begin{figure*}[t]
    \centering
    \includegraphics[width=\linewidth]{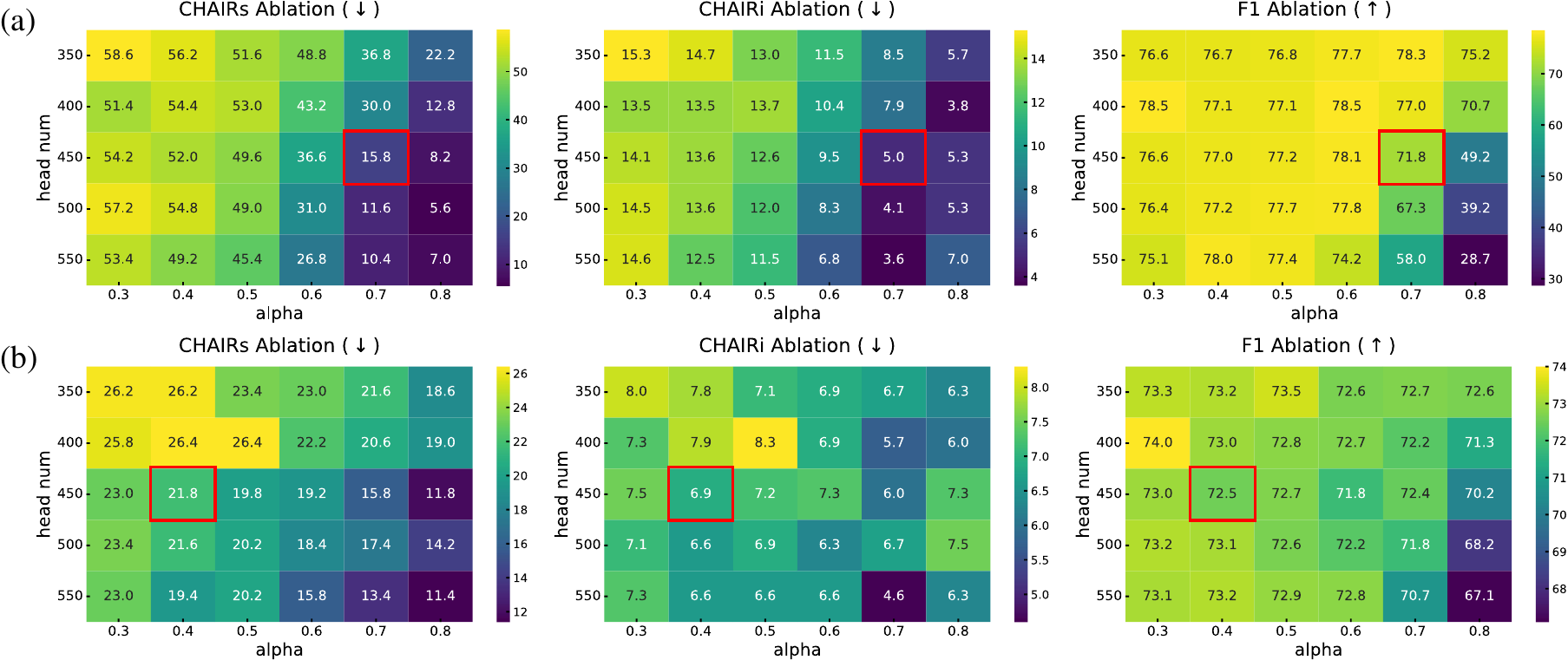}
    \caption{Ablation on hyperparameters $\alpha$ and $K$ for Shikra (a) and MiniGPT4 (b). Red boxes highlight the parameter combinations we used.}
    \label{fig:shikra_minigpt4_abaltion}
\end{figure*}

\section{Supplementary Experiments Results}
\label{app:sup_exp}

\subsection{Performance with Alternative Decoding Strategies}
\label{app:decoding}

In this subsection, we present additional experimental results for the beam search and nucleus sampling decoding strategies, which were omitted from the main text due to space constraints. 
For beam search, we set the beam size to 5, and for nucleus sampling, the temperature is set to 0.5. 
The detailed results are presented in \Cref{tab:main_results_beam} and \Cref{tab:main_results_sample}, respectively. 
Overall, the findings are consistent with the conclusions drawn from the greedy decoding experiments in the main text: our method generally outperforms all baseline approaches across these different strategies.

\subsection{AMBER Results on Additional Models}
\label{app:amber}

In this section, we present the comprehensive experimental results on the AMBER benchmark. 
As shown in \Cref{tab:amber_combined}, the findings align with the conclusions in the main text: \textbf{HAVAE} consistently demonstrates clear superiority over baseline approaches, validating its robust generalization capabilities.

\subsection{Supplementary Ablation Results}
\label{app:albation}

In this subsection, we present the supplementary hyperparameter ablation studies for the remaining models, which were omitted from the main text due to space constraints. 
Specifically, \Cref{fig:shikra_minigpt4_abaltion} illustrates the results for Shikra and MiniGPT4. 
These findings align with the trade-offs discussed in the main paper, further verifying the criticality of precise parameter tuning in long-context scenarios.

\begin{figure*}[t]
    \centering
    \includegraphics[width=\linewidth]{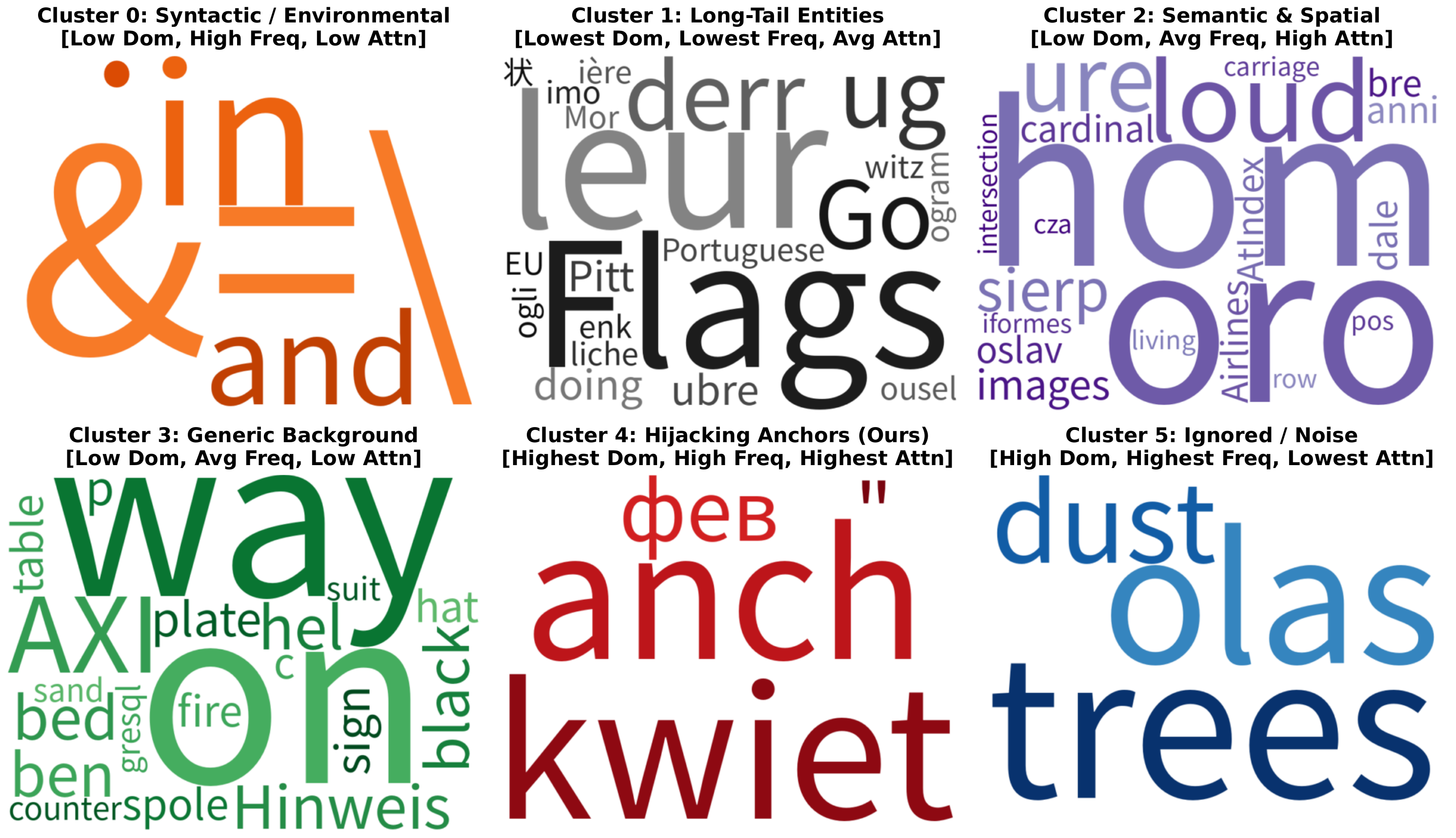} 
    \caption{\textbf{Word cloud visualization of token clusters identified via K-Means clustering on LLaVA-1.5 7B.} 
    Tokens are grouped based on their behavioral signatures in the feature space (Dominance, Frequency, Attention). 
    \textbf{Cluster 4 (Hijacking Anchors)} distinctively isolates the rigid, high-attention artifacts (e.g., \texttt{kwiet}, \texttt{"}) identified by our HABI method, separating them from the active semantic content in \textbf{Cluster 2} and the ignored environmental noise in \textbf{Cluster 5}. 
    This qualitative separation confirms that our quantitative metrics align with the intrinsic semantic structure of the model's vocabulary.}
    \label{fig:cluster}
\end{figure*}

\begin{figure*}[t]
    \centering

    \begin{subfigure}{\linewidth}
        \centering
        \includegraphics[width=\linewidth]{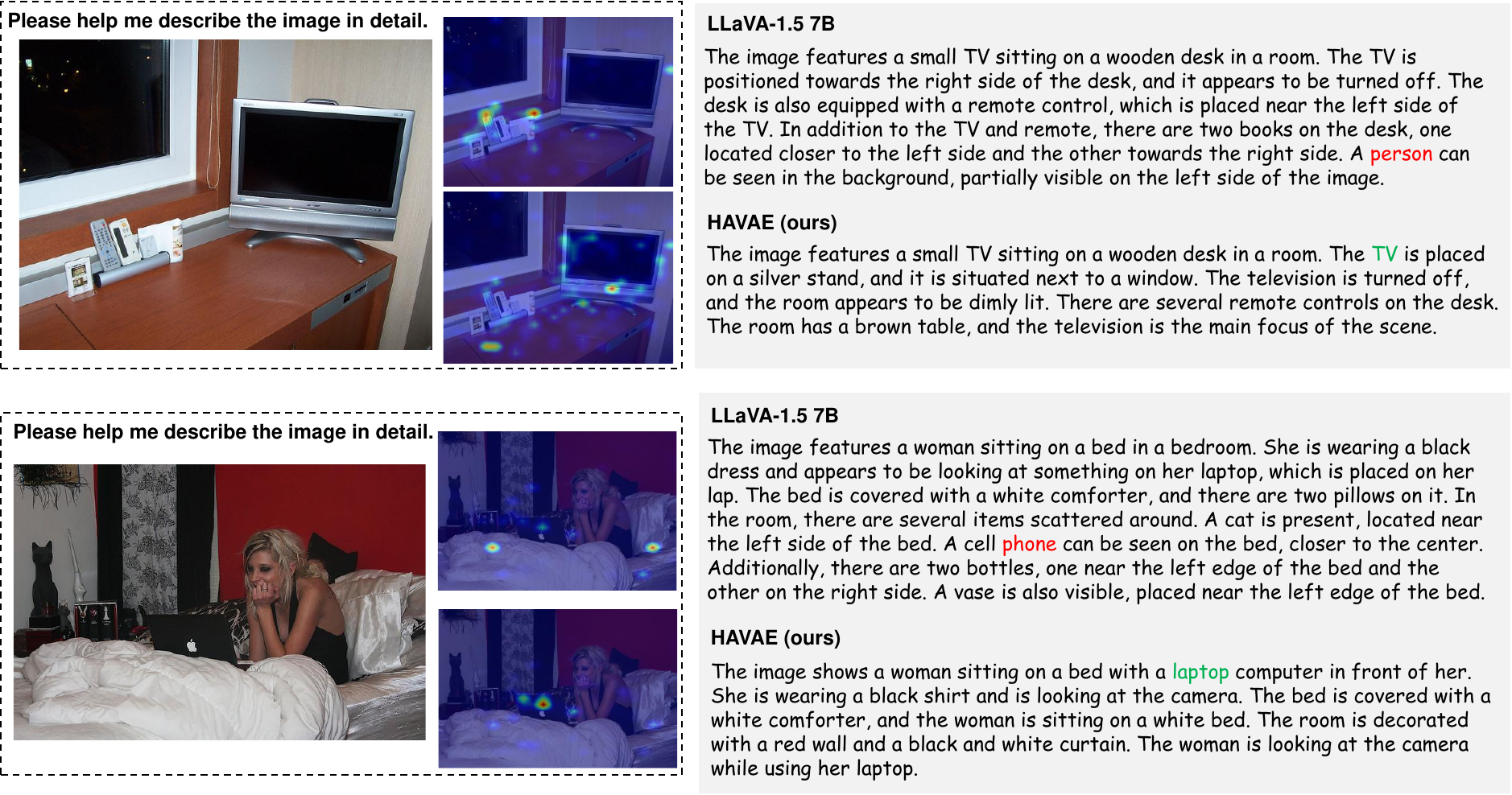}
        \label{fig:case1}
    \end{subfigure}
    
    \vspace{-3mm} 

    \begin{subfigure}{\linewidth}
        \centering
        \includegraphics[width=\linewidth]{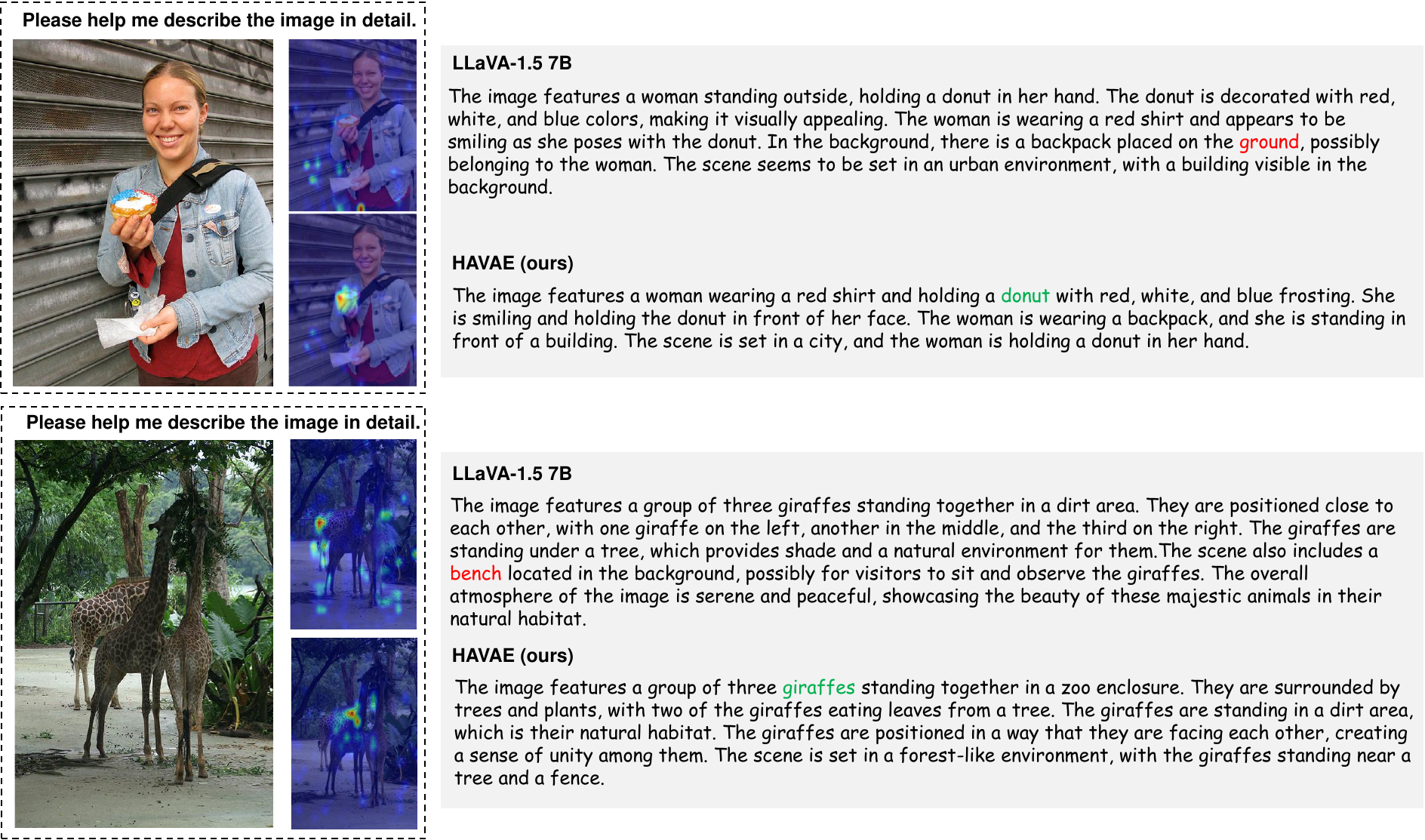}
        \label{fig:case2}
    \end{subfigure}
    
    \caption{
        Additional qualitative comparison of attention maps. For each case, we contrast the attention map for a hallucinated token from the baseline model (top row) with a corresponding real object token from our HAVAE-enhanced model (bottom row), demonstrating HAVAE's ability to refocus attention on salient objects.
    }
    \label{fig:cases_combined}
\end{figure*}

\begin{figure*}[t]
    \centering 
    
    \begin{subfigure}{1\linewidth}
        \centering
        \includegraphics[width=\linewidth]{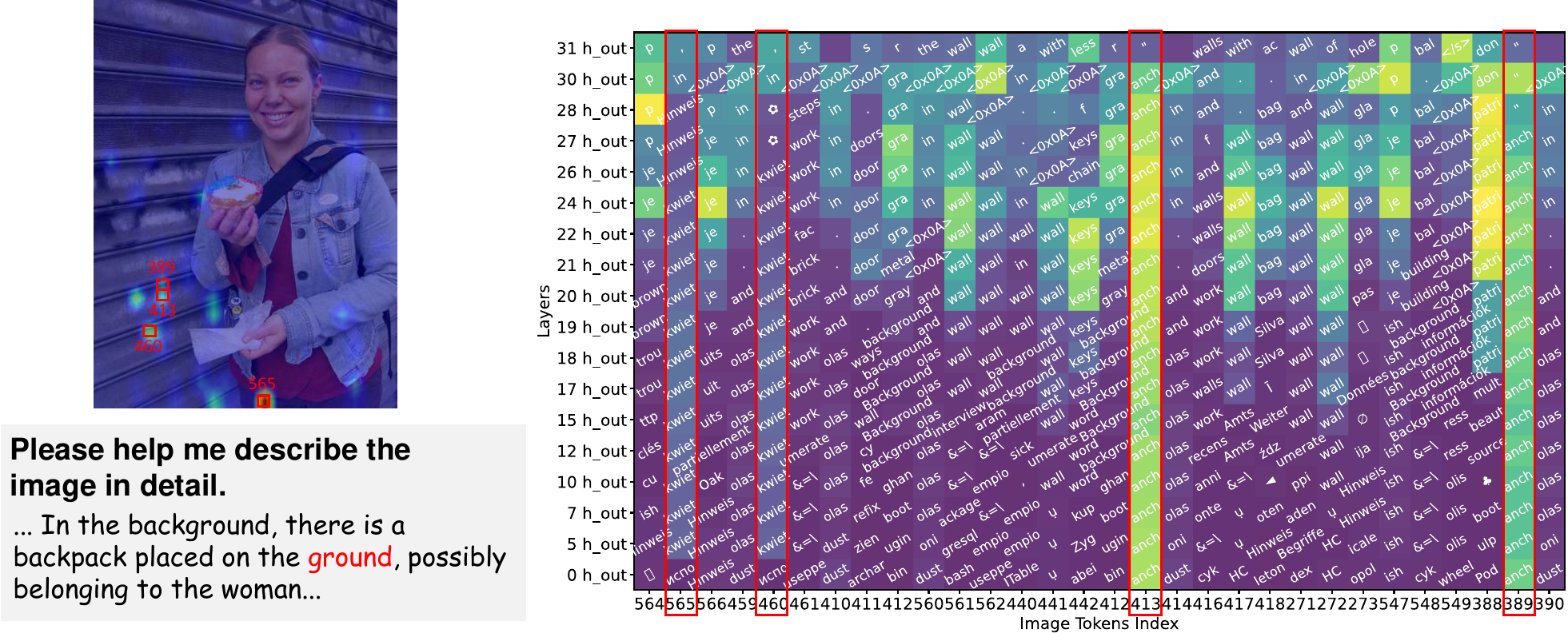}
    \end{subfigure}
    
    \vspace{1em} 
    
    \begin{subfigure}{1\linewidth}
        \centering
        \includegraphics[width=\linewidth]{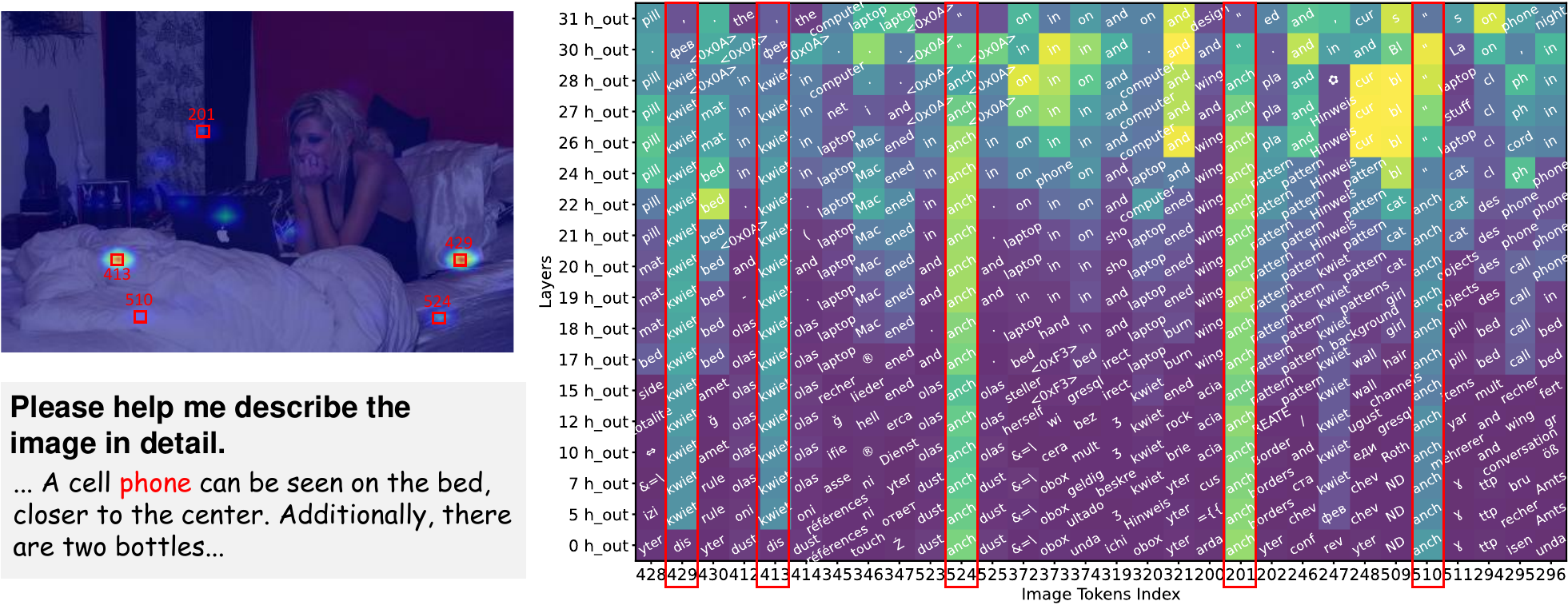}
    \end{subfigure}

    \begin{subfigure}{1\linewidth}
        \centering
        \includegraphics[width=\linewidth]{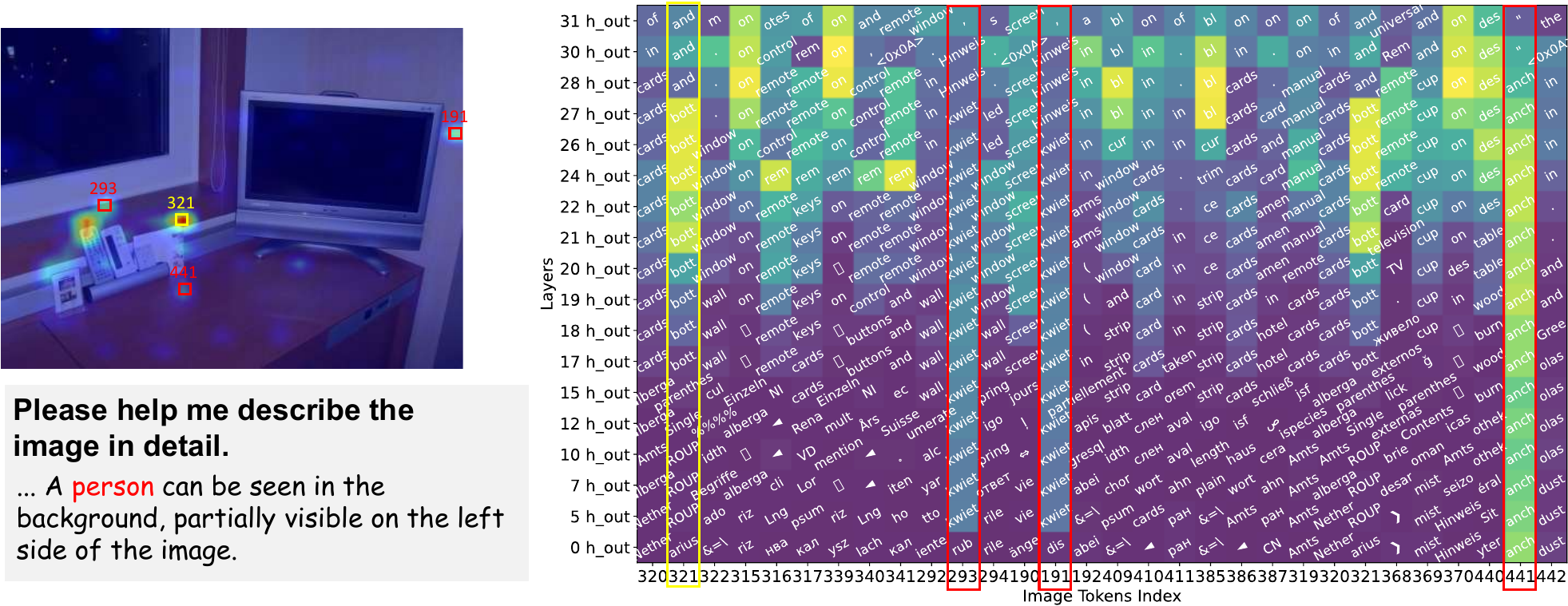}
    \end{subfigure}
    
    \caption{Logit lens visualization for three case studies from \Cref{app:case}. These plots provide further qualitative evidence that the model allocates disproportionately high attention to the identified \textbf{Inert Tokens} during hallucination.}
    \label{fig:vfi_cases}
\end{figure*}

\section{Supplementary Experiments and Analysis}
\label{app:analysis}

\subsection{Qualitative Validation via Clustering and Visualization}
\label{app:cluster_analysis}

In the main text, we utilized a composite quantitative metric (Hijacking Score) combined with IQR thresholding to identify \textbf{Hijacking Anchors}.

To corroborate the distinct properties of these Anchors, we perform an unsupervised K-Means clustering analysis on the visual tokens $\mathcal{V}$, using LLaVA-1.5 7B. 
Each visual token is represented in the feature space defined by (Dominance, Frequency, Attention). 
Subsequent to clustering, we assign each unique \textbf{vocabulary item} (Anchor) to a specific cluster based on the majority distribution of its corresponding visual tokens.

The corresponding word cloud visualizations for each cluster—comprising the Anchors assigned to them—are presented in~\Cref{fig:cluster}.
The results validate our approach by showing that the feature space naturally disentangles vocabulary items into distinct functional roles~\citep{du2024pcb}.

\subsubsection{High-Attention Manifold: Anchors vs. Content}
The critical validation lies in the contrast between \textbf{Cluster 4} and \textbf{Cluster 2}. Both clusters are composed of Anchors associated with high-attention visual tokens, but they exhibit opposite representational stabilities.

\paragraph{Cluster 4: Hijacking Anchors (Ours).}
\textit{Signature: [Highest Dom, High Freq, Highest Attn]. Anchors: \texttt{kwiet}, \texttt{anch}, \texttt{"}, \texttt{\textcyr{фев}}.} \\
This cluster corresponds precisely to our identified \textbf{Hijacking Anchors}. The visual tokens mapped to these Anchors peak in both Dominance and Attention. This unique combination confirms their role as \textbf{``Active Sinks''}: they actively monopolize attention yet remain representationally rigid. Composed largely of meaningless artifacts and delimiters, these Anchors serve as fixed decoding endpoints rather than contextualized meanings.

\paragraph{Cluster 2: Semantic \& Spatial Concepts.}
\textit{Signature: [Low Dom, Avg Freq, High Attn]. Anchors: \texttt{intersection}, \texttt{row}, \texttt{living}, \texttt{Airlines}.} \\
While sharing the high-attention profile of Cluster 4, the visual tokens associated with these Anchors exhibit \textbf{Low Dominance}. This signature indicates \textbf{active processing}: the representations undergo significant shifts as the model extracts meaning. The presence of complex spatial and semantic concepts (\texttt{intersection}, \texttt{row}) confirms that the model correctly distinguishes meaningful content from rigid Hijacking Anchors in the feature space.

\subsubsection{High-Frequency Manifold: Noise vs. Substrate}
This group differentiates frequent Anchors based on the attention levels of their corresponding visual tokens.

\paragraph{Cluster 5: Ignored / Noise Artifacts.}
\textit{Signature: [High Dom, Highest Freq, Lowest Attn]. Anchors: \texttt{trees}, \texttt{dust}, \texttt{olas}.} \\
This cluster captures Anchors associated with environmental noise. The corresponding visual tokens exhibit High Frequency but Lowest Attention. Their High Dominance here reflects \textbf{``Passive Stability''}—these tokens remain static because the model ignores them (bypassing computation), which is mechanically distinct from the ``Active Rigidity'' of Cluster 4.

\paragraph{Cluster 0: Syntactic / Environmental Markers.}
\textit{Signature: [Low Dom, High Freq, Low Attn]. Anchors: \texttt{.}, \texttt{and}, \texttt{in}, \texttt{\&=\textbackslash}.} \\
Unlike Cluster 5, the corresponding visual tokens in this cluster show \textbf{Low Dominance}. This suggests that these tokens (decoding to syntactic markers like \texttt{.}, \texttt{and}, \texttt{in}) actively participate in local structure formation ("syntactic glue") without capturing the global attention reserved for Hijacking Anchors or semantic targets.

\subsubsection{Long-Tail \& Generic Groups}
The final category encompasses the broad spectrum of standard semantic content. It distinguishes between specific, context-dependent entities and generic background elements based on their frequency and specificity.
\paragraph{Cluster 1: Long-Tail Entities.}
\textit{Signature: [Lowest Dom, Lowest Freq, Avg Attn]. Anchors: \texttt{EU}, \texttt{Portuguese}, \texttt{Flags}.} \\
These Anchors represent rare entities and proper nouns. The \textbf{Lowest Dominance} of their corresponding visual tokens indicates \textbf{extensive semantic evolution}. Unlike common concepts, these rare entities require significant layer-wise refinement and context aggregation to be correctly resolved, preventing their representations from stabilizing early.

\paragraph{Cluster 3: Generic Background Concepts.}
\textit{Signature: [Low Dom, Avg Freq, Low Attn]. Anchors: \texttt{on}, \texttt{table}, \texttt{plate}, \texttt{way}.} \\
These Anchors represent the statistical median. While containing common visual objects (\texttt{table}, \texttt{plate}) and prepositions (\texttt{on}), the Low Attention of their corresponding visual tokens suggests they often serve as common background elements rather than primary attentional focal points.

\begin{table}[t]
\centering
\resizebox{\linewidth}{!}{%
\begin{tabular}{l|ccc|cc}
\toprule
\multirow{2}{*}{\textbf{Method}} & \multicolumn{3}{c}{\textbf{CHAIR}} & \multicolumn{2}{c}{\textbf{POPE}} \\
\cmidrule(lr){2-4} \cmidrule(lr){5-6}
 & \textbf{CHAIR$_s$ $\downarrow$} & \textbf{CHAIR$_i$ $\downarrow$} & \textbf{F1 $\uparrow$} & \textbf{Acc. $\uparrow$} & \textbf{F1 $\uparrow$} \\
\midrule
\textbf{HAVAE} & \textbf{18.2} & \textbf{3.7} & \cellcolor{gray!15}76.7 & \textbf{86.1} & \textbf{86.2} \\
w/o GT & 18.6 & 5.0 & \cellcolor{gray!15}78.1 & 85.9 & 86.0 \\ 
\bottomrule
\end{tabular}
}
\caption{Ablation study on the necessity of ground-truth (GT) annotations. ``w/o GT'' denotes the variant where head selection relies solely on generated objects. Results confirm HAVAE's robustness to the absence of external supervision.}
\label{tab:ablation_coco}
\end{table}
\subsection{Robustness to the Absence of Ground-Truth Annotations}
\label{sec:ablation_no_gt}

To validate the applicability of our method in scenarios lacking ground-truth (GT) object annotations, we conducted a comparative ablation study. 
Specifically, we modified the head selection criterion in equation \eqref{eq:mean_nhar} by substituting the ground-truth object set $\mathcal{O}_{\mathrm{real}}$ with the entire set of generated objects (i.e., the union $\mathcal{O}_{\mathrm{real}} \cup \mathcal{O}_{\mathrm{hal}}$). 
For this setting, we adjusted the hyperparameters to $\alpha=0.5$ and the number of selected heads to 550, while keeping all other configurations unchanged. 
The comparative results on the CHAIR and POPE benchmarks are presented in \Cref{tab:ablation_coco}.
The empirical evidence demonstrates that HAVAE maintains superior performance even without reliance on external ground-truth annotations, thereby confirming its robustness and potential for broader scalability.

\section{Additional Case Studies}
\label{app:case}

In this section, we present several additional case studies to qualitatively demonstrate how our HAVAE method concentrates attention on key visual content and effectively mitigates hallucination. \Cref{fig:cases_combined} visualizes four distinct examples. For each example, we contrast the attention map for a hallucinated object token generated by the baseline model with the attention map for a corresponding real object token from our HAVAE-enhanced model. It is important to note a key methodological difference in these visualizations: the baseline maps represent the average attention across \textbf{all} heads, whereas the maps for our method show the average attention of the \textbf{450 heads selected by HAVAE}.

Additionally, for three of these examples, we provide logit lens visualizations in \Cref{fig:vfi_cases} (following the approach in \Cref{fig:intro}) to further illustrate the dynamics of \textbf{Vocabulary Hijacking}. These visualizations confirm that the model allocates disproportionately high attention to specific \textbf{Hijacking Anchors} when a hallucination is produced.

From these examples, we can clearly observe a consistent pattern: when a hallucination occurs, the baseline model's attention is often hijacked by irrelevant \textbf{Inert Tokens}. The effect of HAVAE is visually evident, as it successfully refocuses the model's attention onto the grounded target object.

\subsection{Qualitative Error Analysis}
\label{app:error_analysis}

While our method successfully mitigates hallucinations driven by Inert Tokens, we provide a qualitative error analysis to discuss a limitation of our current observational framework. 

Revisiting the third case presented in \Cref{fig:vfi_cases}, we observe an intriguing anomaly: the $321^{\text{st}}$ visual token captures an anomalously high proportion of the attention weights, despite being completely semantically irrelevant to the currently generated text token. Interestingly, however, this specific token does \textit{not} exhibit the characteristic Vocabulary Hijacking phenomenon. 

This counter-example indicates that while Vocabulary Hijacking is a highly prevalent and actionable indicator of LVLM hallucinations, it cannot deterministically localize \textit{all} instances of attention misallocation. There remain atypical failure modes where the model assigns incorrect attention without triggering the semantic collapse associated with Inert Tokens, suggesting that complementary mechanisms may be required to comprehensively resolve these diverse attention errors in future work.

\section{Algorithms}
\label{app:algorithm}

For clarity and to facilitate reproducibility, this section provides the detailed algorithmic procedures for our proposed frameworks. We present the \textbf{Hijacking Anchor-Based Identification (HABI)} framework in \Cref{alg:habi} and the \textbf{Hijacking-Aware Visual Attention Enhancement (HAVAE)} framework in \Cref{alg:havae}.

\begin{algorithm*}[t]
\caption{Hijacking Anchor-Based Identification (HABI)}
\label{alg:habi}
\begin{algorithmic}[1]
\REQUIRE LVLM model $M$; Calibration dataset $\mathcal{D}$ (e.g., 500 COCO images).
\ENSURE The set of Inert Tokens $\mathcal{I}_{\text{inert}}$ for a given input image's visual tokens $\mathcal{I}_\text{v}$.

\STATE
\STATE /* \textbf{Phase 1: Global Calibration \& Anchor Discovery} (Corresponds to Sec. \ref{sec:case_setup} \& \ref{sec:vf}) */
\STATE $\mathcal{V} \leftarrow \emptyset$, $AnchorMap \leftarrow \{\}$, $AllScores \leftarrow []$
\FOR{each image $I \in \mathcal{D}$}
    \STATE $\mathcal{I}_v \leftarrow \text{GetVisionTokens}(M, I)$
    \FOR{each vision token $v_i \in \mathcal{I}_v$}
        \STATE $\mathcal{T}_{v_i} \leftarrow \text{LogitLensDecode}(M, v_i)$ \COMMENT{Get Trace}
        \STATE $a_{v_i} \leftarrow \text{MostFrequentToken}(\mathcal{T}_{v_i})$ \COMMENT{Identify Trace Anchor}
        \STATE Calculate metrics: Dominance $\mathcal{D}(v_i)$, Frequency $\mathcal{F}(v_i)$, Attention $\mathcal{A}(v_i)$
        \STATE $S_{\text{hijack}}(v_i) \leftarrow \mathcal{D}(v_i) \cdot \mathcal{F}(v_i) \cdot \mathcal{A}(v_i)$
        \STATE $\text{Append}(AllScores, S_{\text{hijack}}(v_i))$ \COMMENT{Collect individual scores}
        \STATE $\text{Append}(AnchorMap[a_{v_i}], S_{\text{hijack}}(v_i))$ \COMMENT{Group scores by anchor word}
        \STATE $\mathcal{V} \leftarrow \mathcal{V} \cup \{v_i\}$
    \ENDFOR  
\ENDFOR

\STATE
\STATE /* \textbf{Phase 2: Threshold Determination} (Corresponds to Sec. \ref{sec:vf} \& HABI) */
\STATE /* 2.1 Identify Hijacking Anchors */
\STATE $\tau_s \leftarrow Q_3(AllScores) + 1.5 \cdot \text{IQR}(AllScores)$ \COMMENT{Outlier threshold based on global score distribution}
\STATE $\mathcal{A}_{\text{hijack}} \leftarrow \emptyset$
\FOR{each unique anchor $w \in \text{Keys}(AnchorMap)$}
    \IF{$\text{Mean}(AnchorMap[w]) > \tau_s$}
        \STATE $\mathcal{A}_{\text{hijack}} \leftarrow \mathcal{A}_{\text{hijack}} \cup \{w\}$
    \ENDIF
\ENDFOR

\STATE
\STATE /* 2.2 Determine Hijacking Ratio Threshold using Otsu's Method */
\STATE $Ratios \leftarrow []$
\FOR{each $v_i \in \mathcal{V}$ \textbf{if} $v_i \in \text{Top5\%Attention}$}
    \STATE $r_{\text{hijack}}(v_i) \leftarrow \frac{1}{|\mathcal{T}_{v_i}|} \sum_{t \in \mathcal{T}_{v_i}} \mathbb{I}(t \in \mathcal{A}_{\text{hijack}})$
    \STATE $\text{Append}(Ratios, r_{\text{hijack}}(v_i))$
\ENDFOR
\STATE $\tau_r \leftarrow \text{OtsuMethod}(Ratios)$ \COMMENT{Find separation threshold in bimodal distribution}

\STATE
\STATE /* \textbf{Phase 3: HABI Inference Function} (Corresponds to HABI Implementation) */
\STATE \textbf{function} IdentifyInertTokens($\mathcal{I}_\text{v}, M, \mathcal{A}_{\text{hijack}}, \tau_r$)
    \STATE $\mathcal{I}_{\text{inert}} \leftarrow \emptyset$
    \FOR{each vision token $v_i \in \mathcal{I}_\text{v}$}
        \STATE $\mathcal{T}_{v_i} \leftarrow \text{LogitLensDecode}(M, v_i)$
        \STATE $r_{\text{hijack}}(v_i) \leftarrow \frac{1}{|\mathcal{T}_{v_i}|} \sum_{t \in \mathcal{T}_{v_i}} \mathbb{I}(t \in \mathcal{A}_{\text{hijack}})$ \COMMENT{Calculate Hijacking Ratio}
        \IF{$r_{\text{hijack}}(v_i) > \tau_r$}
            \STATE $\mathcal{I}_{\text{inert}} \leftarrow \mathcal{I}_{\text{inert}} \cup \{v_i\}$
        \ENDIF
    \ENDFOR
    \STATE \textbf{return} $\mathcal{I}_{\text{inert}}$
\STATE \textbf{end function}

\end{algorithmic}
\end{algorithm*}

\begin{algorithm*}[t]
\caption{Hijacking-Aware Visual Attention Enhancement (HAVAE) Framework}
\label{alg:havae}
\begin{algorithmic}[1]
\REQUIRE LVLM model $M$; Calibration dataset $\mathcal{D}$; HABI function `IdentifyInertTokens' (from Alg. \ref{alg:habi}); Pre-computed Hijacking Anchors $\mathcal{A}_{\text{hijack}}$, $\tau_r$ (HABI threshold); Hyperparameters $K$ (num heads), $\alpha$ (strength).
\ENSURE A modified inference process with mitigated hallucination.

\STATE
\STATE /* \textbf{Stage 1: Principled Head Selection (Offline)} */
\STATE \textbf{function} SelectCriticalHeads($M, \mathcal{D}, K$)
    \STATE $\mathcal{O}_{\mathrm{real}} \leftarrow \text{CollectRealObjectTokens}(M, \mathcal{D})$ \COMMENT{Correspond to \Cref{sec:case_setup}}
    \STATE $\overline{\text{NHAR}} \leftarrow \text{InitializeMatrix}(L, H, \text{zeros})$
    
    \FOR{each real object token $y_k \in \mathcal{O}_{\mathrm{real}}$}
        \STATE $\mathcal{I}_\text{v} \leftarrow \text{GetCorrespondingVisualTokens}(y_k)$
        \STATE $\mathcal{I} \leftarrow \text{GetFullContext}(y_k)$
        \STATE $\mathcal{I}_{\text{inert}} \leftarrow \text{IdentifyInertTokens}(\mathcal{I}_\text{v}, M, \mathcal{A}_{\text{hijack}}, \tau_r)$ \COMMENT{\textbf{Call HABI function}}
        \STATE $A_{\text{all\_heads}} \leftarrow \text{GetAttentionForAllHeads}(M, y_k)$
        
        \FOR{each head $(\ell, h)$ from $(1,1)$ to $(L,H)$}
            \STATE /* Calculate NHAR for this step */
            \STATE $A_{\ell,h} \leftarrow A_{\text{all\_heads}}[\ell, h]$
            \STATE $Attn_{\text{valid\_visual}} \leftarrow \sum_{v_i \in \mathcal{I}_\text{v} \setminus \mathcal{I}_{\text{inert}}} A_{\ell,h,i}$
            \STATE $Attn_{\text{total}} \leftarrow \sum_{v_i \in \mathcal{I}} A_{\ell,h,i}$ 
            
            \STATE $\text{NHAR\_score} \leftarrow Attn_{\text{valid\_visual}} / Attn_{\text{total}}$ 
            \STATE $\overline{\text{NHAR}}_{\ell,h} \leftarrow \overline{\text{NHAR}}_{\ell,h} + \text{NHAR\_score}$
        \ENDFOR
    \ENDFOR
    
    \STATE $\overline{\text{NHAR}} \leftarrow \overline{\text{NHAR}} / |\mathcal{O}_{\mathrm{real}}|$ \COMMENT{Eq. \ref{eq:mean_nhar}}
    \STATE $\mathcal{H}_{\text{target}} \leftarrow \text{TopK}(\overline{\text{NHAR}}, K)$ \COMMENT{Select critical heads}
    \STATE \textbf{return} $\mathcal{H}_{\text{target}}$
\STATE \textbf{end function}

\STATE
\STATE /* \textbf{Stage 2: Collective Attention Reinforcement (Online)} */
\STATE \textbf{function} HAVAE\_Attention\_Forward($A_{original}, \mathcal{H}_{\text{target}}, \alpha$)
    \STATE \COMMENT{Injects this logic into the model's attention mechanism during inference.}
    \STATE $L, H \leftarrow \text{GetModelDimensions}()$
    \STATE $A_{enhanced} \leftarrow A_{original}$

    \FOR{$\ell = 1$ to $L$}
        \STATE /* Calculate Layer-wise Mean Attention (Eq. \ref{eq:attention_enhancement}) */
        \STATE $MeanAttn_\ell \leftarrow \text{Zeros}(\text{size}=\text{num\_visual\_tokens})$
        \FOR{$h' = 1$ to $H$}
            \STATE $MeanAttn_\ell \leftarrow MeanAttn_\ell + |A_{original}[\ell, h', \text{visual\_indices}]|$
        \ENDFOR
        \STATE $MeanAttn_\ell \leftarrow MeanAttn_\ell / H$
        
        \STATE /* Apply reinforcement to target heads */
        \FOR{$h = 1$ to $H$}
            \IF{$(\ell, h) \in \mathcal{H}_{\text{target}}$}
                \STATE $A_{enhanced}[\ell, h, \text{visual\_indices}] \leftarrow A_{original}[\ell, h, \text{visual\_indices}] + \alpha \cdot MeanAttn_\ell$
            \ENDIF
        \ENDFOR
    \ENDFOR
    \STATE \textbf{return} $A_{enhanced}$
\STATE \textbf{end function}

\end{algorithmic}
\end{algorithm*}

\end{document}